\documentclass{llncs}

\usepackage{graphicx}

\usepackage{makecell} % for line breaks in tables
\usepackage{amssymb}
\usepackage[caption=false,subrefformat=parens, labelformat=parens, font=footnotesize,margin=20pt]{subfig}
\usepackage{stfloats}
\usepackage[strings]{underscore}
\usepackage{color, colortbl}

\begin{document}

\title{Exploring the Design Space of Aesthetics with the Repertory Grid Technique}
\author{David Baum\inst{1}\orcidID{0000-0002-0213-0277}}

\institute{Leipzig University, Grimmaische Straße 12, 04109 Leipzig, Germany}

\maketitle

\begin{abstract}
By optimizing aesthetics, graph diagrams can be generated that are easier to read and understand.
However, the challenge lies in identifying suitable aesthetics.
%This process is currently mainly based on intuition and has some shortcomings.
We present a novel approach based on repertory grids to explore the design space of aesthetics systematically.
%Aesthetics can thus be obtained more efficiently, reproducibly, and less subjective.
% To evaluate our approach, we applied it to graph diagrams.
% To establish a ground truth, we conducted a literature study to find all published graph aesthetics.
We applied our approach with three independent groups of participants to systematically identify graph aesthetics.
In all three cases, we were able to reproduce the aesthetics with positively evaluated influence on readability without any prior knowledge.
We also applied our approach to two- and three-dimensional domain-specific software visualizations to demonstrate its versatility.
In this case, we were also able to acquire several aesthetics that are relevant for perceiving the visualization.
% These aesthetics will be used in future work to increase the readability and comprehensibility of these diagrams.
\end{abstract}  

%%%%%%%%%%%%%%%%%%%%%%%%%%%%%%%%%%%%%%%%%%%%%%%%%%%%%%%%%%%%%%%%%%%%%%%%%%%%%%%%%%%%%%%%%%%%%%%%%%%%%%%%%%%%%%%%%%%%%%%%%%%%%%%%%%%%%%%%%%%%%%%%%%%%%%%%%%%%%%%%%%%%%%%%%%%%%%%%%%%%%%%%%%%%%%%%%%%%
\section{Introduction}
%%%%%%%%%%%%%%%%%%%%%%%%%%%%%%%%%%%%%%%%%%%%%%%%%%%%%%%%%%%%%%%%%%%%%%%%%%%%%%%%%%%%%%%%%%%%%%%%%%%%%%%%%%%%%%%%%%%%%%%%%%%%%%%%%%%%%%%%%%%%%%%%%%%%%%%%%%%%%%%%%%%%%%%%%%%%%%%%%%%%%%%%%%%%%%%%%%%%

Making visualizations easier to read and to understand is a challenging task and has been researched for decades~\cite{Battista1999}.
Aesthetics are a suitable method to address this problem~\cite{Purchase1997}.
They represent heuristics to predict human perception of the visualization. 
Aesthetics are visual metrics that must be both objectively measurable and perceptible to the observer~\cite{Baum2015}.
They are independent of the semantic context of a visualization and refer only to visual properties.

For graph layouts consisting of nodes and edges, aesthetics are well researched.
Typical aesthetics are, e.g., edge crossings and cutting angles of edges~\cite{Purchase1997}.
These criteria are used as optimization goals, e.g., minimizing the number of edge crossings or maximizing the average cutting angle to generate perceivable and comprehensible graph layouts.
Aesthetics have been adapted to other visualizations, e.g., different sorts of diagrams~\cite{Purchase2001,Effinger2010} as well as complex graphical user interfaces such as websites~\cite{Pajusalu2012}.
Each type of visualization has its own aesthetics.
Therefore, the state of the art research process has to be repeated for every type of visualization.
The process is always similar and comprises the following steps. 

\begin{enumerate}
 \item \textbf{Define one or multiple aesthetics.}
 Every aesthetic must be measurable.
 There is no established way to derive aesthetics.
 Many aesthetics are only chosen because they seem to be plausible, so this step is subjective.

 \item \textbf{Evaluate impact of proposed aesthetics empirically.}
 In this step, participants solve tasks using different visualizations, measuring error rate and time to complete the task.
 It is necessary to be able to trace possible differences in solving the tasks back to different aesthetics.
 This can be achieved, for example, by changing one aesthetic while keeping all others approximately constant.
 This is often only possible to a limited extent due to dependencies between different aesthetics.
 
 \item \textbf{Implement layout algorithm.}
 To make the positively evaluated aesthetics usable in practice, it is necessary to provide a suitable layout algorithm.
 It should have a reasonable runtime behavior and take care of conflicting optimization goals.
\end{enumerate}

The whole process is iterative.
Depending on the procedure, step 3 might be performed before step 2. % zur Not streichen
When new aesthetics are defined, the subsequent steps have to be repeated accordingly.
However, this approach leads to significant problems.
Without being aware of all relevant aesthetics, interactions between them cannot be considered.
% it is not possible to keep all but one constant.
Unknown but relevant aesthetics might distort the outcome of empirical evaluations significantly~\cite{Huang2010}.
% As a result, many of the optimization layout algorithms developed to date only take one or two aesthetics into account.
% There is no layout algorithm that considers all prevalent aesthetics.
In addition, some aesthetics are not obvious, especially for complex visualizations with many different visual primitives.
Hence, there is a risk that important aesthetics may be overlooked.
The whole process is very tedious because aesthetics are also defined and examined that have no measurable effect on readability.

In this paper, we want to improve the identification of aesthetics by making the process more reproducible and less based on the researcher's intuition.
We use a novel approach based on the repertory grid technique (RGT).
This is an interview technique that triggers the participants' creativity to describe verbally the differences between certain elements.
These descriptions then serve as a basis for the definition of aesthetics.
Therefore, more relevant aesthetics are known when it comes to conducting the evaluations.
This will simplify the outlined research process and help to overcome the mentioned problems.
% We first review related work, followed by a description of our implementation of the repertory grid technique (RGT).
% We conducted three empirical studies to evaluate our approach and show its versatility.
% In the first study we apply our approach to node-link diagrams and reproduce already published aesthetics.
% Then we present two studies, where repertory grids have been used to derive aesthetics from complex two- and three-dimensional domain specific visualizations.
% Finally, we discuss the results, followed by a brief conclusion.

%%%%%%%%%%%%%%%%%%%%%%%%%%%%%%%%%%%%%%%%%%%%%%%%%%%%%%%%%%%%%%%%%%%%%%%%%%%%%%%%%%%%%%%%%%%%%%%%%%%%%%%%%%%%%%%%%%%%%%%%%%%%%%%%%%%%%%%%%%%%%%%%%%%%%%%%%%%%%%%%%%%%%%%%%%%%%%%%%%%%%%%%%%%%%%%%%%%%%
\section{Related Work}
%%%%%%%%%%%%%%%%%%%%%%%%%%%%%%%%%%%%%%%%%%%%%%%%%%%%%%%%%%%%%%%%%%%%%%%%%%%%%%%%%%%%%%%%%%%%%%%%%%%%%%%%%%%%%%%%%%%%%%%%%%%%%%%%%%%%%%%%%%%%%%%%%%%%%%%%%%%%%%%%%%%%%%%%%%%%%%%%%%%%%%%%%%%%%%%%%%%%

% The research on the design space of visualizations is an essential component of information visualization~\cite{Schulz2013}.
Several models and guidelines exist for designing and evaluating visualizations~\cite{Munzner2009,McKenna2014,Meyer2015}.
% An elaborated framework for this was established by Munzner et al.~\cite{Munzner2009}.
% The repertory grid technique has been used to explore the design space of visualizations before \cite{Kurzhals2018,Baum2015}.
% Kurzhals et al. \cite{Kurzhals2018} use an approach to explore the design space of visualizations based on the repertory grid technique.
However, these process models do not use any aesthetics.
The only framework known to us that takes aesthetics into account is~\cite{Lau2007}.
It assumes that aesthetics and its effects are already known.
Most aesthetics are selected based on intuition without giving an explicit rationale.
Bennett et al.~\cite{Bennett2007} justify established aesthetics with Gestalt principles.
However, they do not show how new aesthetics can be derived from Gestalt principles.

We are only aware of one approach to improve the iterative process by making it less subjective and more efficient: drawings \cite{Papadopoulos2013}.
% Most aesthetics were simply set up.
% They were mostly derived from the design principles and found to be plausible, but in some cases not at all or only years later empirically evaluated.
% It turned out that many of the aesthetics used had no significant influence on the legibility and comprehensibility.
% Two strategies have emerged: Drawings and Eye Tracking.
The participants are asked to draw visualizations with a given structure, often node-link diagrams.
Subsequently, it is examined by statistical means which aesthetics the respondent applied to their drawing.
% For example, it can be checked whether the subject has drawn straight or bent edges, whether they has tried to avoid overlapping edges, and so on \cite{Papadopoulos2013}.
% Subsequently, the frequently used aesthetics in particular are examined further, where, on the other hand, the unused aesthetics seem to be unimportant for the subject.
Drawings can help to some extent to weigh aesthetics or identify any irrelevant aesthetics.
However, the capabilities of this approach to explore the aesthetics design space are limited.
This approach requires well-defined aesthetics to check if they have been used by the subject or not.
Also, drawings will not work for complex or three-dimensional visualizations, since most participants will be unable to express their mental model adequate in a drawing of such visualizations.

We see drawings as a step towards improving the described research process.
Nevertheless, some problems remain unsolved, which we address within this paper.
Our approach is based on the RGT.
We are not aware that this method has already been used in the context of aesthetics.
% which has already been used in other work in the context of information visualization.
% There are some works that discuss the repertory grid technique in the context of information visualization.
% Kurzhals et al.~\cite{Kurzhals2018} used repertory grids to explore the design space of visualizations without considering aesthetics.
% In other works the RGT is used to investigate mental maps~\cite{Mayr2016}, to analyze the the perceived characteristics of visualizations with spatial information~\cite{Meng2004}, and to evaluate user experience~\cite{Vermeeren2010,Aziz2016}
In our previous work~\cite{Baum2015} we used RGT to identify neglected and overemphasized information in visualizations.

%%%%%%%%%%%%%%%%%%%%%%%%%%%%%%%%%%%%%%%%%%%%%%%%%%%%%%%%%%%%%%%%%%%%%%%%%%%%%%%%%%%%%%%%%%%%%%%%%%%%%%%%%%%%%%%%%%%%%%%%%%%%%%%%%%%%%%%%%%%%%%%%%%%%%%%%%%%%%%%%%%%%%%%%%%%%%%%%%%%%%%%%%%%%%%%%%%%%%
\section{Repertory Grid Technique}
%%%%%%%%%%%%%%%%%%%%%%%%%%%%%%%%%%%%%%%%%%%%%%%%%%%%%%%%%%%%%%%%%%%%%%%%%%%%%%%%%%%%%%%%%%%%%%%%%%%%%%%%%%%%%%%%%%%%%%%%%%%%%%%%%%%%%%%%%%%%%%%%%%%%%%%%%%%%%%%%%%%%%%%%%%%%%%%%%%%%%%%%%%%%%%%%%%%%

The RGT is an empirical and qualitative research method. % that originates in psychology. 
% It was already adopted to other research fields such as marketing and software engineering \cite{Khamisani2006,Edwards2009}.
% It has been used for evaluating user experience~\cite{Vermeeren2010}. %and for exploring the design space of visualizations~\cite{Kurzhals2018,Baum2015}.
Its basic assumption is that everybody describes and evaluates elements based on a large set of personal constructs that can be expressed by using \emph{bipolar constructs}~\cite[p.~15]{Fransella2003}.
Elements are for example objects, persons, experiences, or even products.
A construct is defined as ``a way in which two or more things are alike and thereby different from a third or more things''~\cite[p.~61]{Kelly1991}.
These constructs consist of two opposite poles, e.g., ``clear'' and ``confusing'' as well as a construct continuum in between, i.e., different degrees of clarity.
The RGT is an approach to make these constructs explicit and visible.
The process is reproducible and facilitates the structured exploration of an unknown domain.
To apply the RGT, multiple design decisions have to be made, e.g., how elements and constructs are selected. 
% In the following we do not discuss all possibilities, but the research design that corresponds to our research questions, hence, variants such as constructs provided by the researcher are not reasonable for exploring design spaces.
In the following, we will discuss the research design that corresponds to our research questions.
We will not discuss variants that are not reasonable for exploring design spaces such as constructs provided by the researcher.

\subsection{Element selection}

Every interview is done with the same set of elements.
They are selected by the researcher and should represent as much breadth of the domain as possible.
The RGT helps to recognize differences between those elements.
Something that all elements have in common will most likely not be taken into account by the participants.
% If a property is the same in all elements, it is very likely that it will not be taken into account by the participants.
For example, if all visualizations only consist of black entities, no constructs for color mapping can be expected.
The constructs obtained in this way are still valid, but it is possible that they only describe a subset of the domain.
This threat can be reduced by asking the subject to provide additional elements that differ from those given~\cite{Bell1991}.
Placeholder elements such as ``ideal visualization'' or ``worst visualization'' can also be used to ensure adequate coverage of the domain~\cite{Fransella2005b}.
The elements can be based on real or artificially generated data.

\subsection{Construct elicitation}

The constructs are not predefined. 
It will be investigated which constructs the participants use to describe the elements shown to them.
For this purpose, three elements are randomly selected and presented at once to the participant (cf.~\ref{fig:grid_ui}.
The participant has to answer the following question:
``How are any two of these alike in some way?'', complemented by ``What is the opposite of that?'' \cite{Fransella2005}.
The answers to both questions are the respective poles.
For example, a participant might describe those visualizations with two bipolar constructs, ``helpful – unhelpful'' and ``ugly – beautiful''.
For them, these are the relevant attributes in which the two visualizations differ.
Constructs differ in their level of abstraction.
Some constructs are abstract, e.g., ``ugly – beautiful'', others are very concrete, e.g., ``no edge crossings – many edge crossings''.
Abstract constructs are less helpful for our research question since they are subjective and hard to measure.
% It might happen that these abstract constructs lead to further constructs if they are investigated in depth.
These abstract constructs might lead to furtheer constructs if they are investigated in depth.
It is not uncommon that a construct implies another construct.
They only vary in their level of abstraction.
The process of using a construct to attain a more concrete construct is called \emph{laddering} and is a common part of the repertory grid interview \cite{Bell1991,Fransella2005b}. 
This can be done by asking ``Why does this visualization appear more beautiful to you?''.
For example, the answer could lead to the construct ``symmetrical – asymmetrical''. % if the subject perceives symmetrical visualizations as more beautiful.
The whole procedure is repeated by using other randomly selected elements as long as the participant creates new constructs to distinguish between the elements.
It is not feasible to use all possible combinations during the interview, hence a reasonable stop criterion is necessary. %, e.g., a fixed number of triads or a dynamic stop criterion.
We advise stopping the interview when three times in a row the participant did not use any new constructs.
This will lead to enough constructs and does not prolong the interview unnecessarily.
During the interview, the participants have no access to any constructs they used before. 
% It is advisable to point out to the participants that for each triad they should use the constructs that are most meaningful to them. 
Otherwise, participants may try to avoid repetition or use synonyms to find as many constructs as possible.
Further, there is no restriction on how many constructs may be named.

The interviewer must understand what the participant describes with a construct.
For this reason, informal communication between both persons is a regular and intended part of the RGT.
This may include further explanations by the participant, showing examples or simple drawings.
% A pretest is always advisable to detect unexpected difficulties.
The RGT demands high standards of the interviewer and the research design.
The interviewer should be familiar with the established guidelines for conducting the interviews.
We have mainly followed the recommendation of Kurzhals et al.~\cite{Kurzhals2018} and Fransella~\cite{Fransella2005b}.

Normally, a repertory grid interview also includes the creation of the name-giving grids.
The participant evaluates for each element and each construct which pole is more appropriate.
For us, however, this information is of little value as we are interested in the constructs used.
For this reason, we have skipped this step.

\subsection{Analysis}

The output of the interview is a list of constructs, that has to be further analyzed.
Some constructs will represent aesthetics directly, but many constructs are not interesting to us.
This is expected and cannot be avoided.
Kurzhals et al. propose the following categorization to analyze the constructs of repertory grid interviews to explore the design space~\cite{Kurzhals2018}:
\begin{itemize}
 \item \textbf{Visual Mapping} 
    This category covers all constructs, that refer to the use of visual primitives (e.g., straight edges – bent edges) and color mapping.
 \item \textbf{Composition}
    This category consists of constructs that refer to the composition of visualization elements, i.e., layout, alignment, and visual density.
 \item \textbf{Data-related}
    Constructs are data-related and therefore belong to the third category, if they depend on the underlying data, such as ``few nodes – many nodes''.
 \item \textbf{Visual experience}
    The last category describes the hedonistic qualities of visualizations, such as ``ugly – beautiful''~\cite{Hassenzahl2014}.
\end{itemize}

For our research question, only the first two categories are interesting since they represent aesthetics.
Data-related constructs do not describe the properties of the visualization but of the underlying data.
Constructs of the last category are often vague and used as a starting point for laddering during the interview.
In the process, more concrete constructs can be revealed that refer to visual mapping or composition.
The last step is to reformulate the constructs as aesthetics.
``straight edges – bend edges'' becomes ``edge curve'' and so on.
This step is straight forward and should not cause any problems.
If ambiguities should arise here, the laddering was not sufficiently performed.
The final result is a comprehensive list of aesthetics.
The method of extraction ensures that all aesthetics are perceivable for human beings.
However, there is no guarantee that all of them will have a significant influence on the readability of the visualization.
 
%%%%%%%%%%%%%%%%%%%%%%%%%%%%%%%%%%%%%%%%%%%%%%%%%%%%%%%%%%%%%%%%%%%%%%%%%%%%%%%%%%%%%%%%%%%%%%%%%%%%%%%%%%%%%%%%%%%%%%%%%%%%%%%%%%%%%%%%%%%%%%%%%%%%%%%%%%%%%%%%%%%%%%%%%%%%%%%%%%%%%%%%%%%%%%%%%%%%
\section{Evaluation}
%%%%%%%%%%%%%%%%%%%%%%%%%%%%%%%%%%%%%%%%%%%%%%%%%%%%%%%%%%%%%%%%%%%%%%%%%%%%%%%%%%%%%%%%%%%%%%%%%%%%%%%%%%%%%%%%%%%%%%%%%%%%%%%%%%%%%%%%%%%%%%%%%%%%%%%%%%%%%%%%%%%%%%%%%%%%%%%%%%%%%%%%%%%%%%%%%%%%

% Graph aesthetics are well-researched for several decades.
Many graph aesthetics have been proposed, and some of them have been evaluated in empirical studies~\cite{Chen2005}.
We define positively evaluated aesthetics as aesthetics for which a significant influence on readability has already been empirically demonstrated.
We applied the RGT to the domain of graph visualization to check the following hypotheses:

\begin{itemize}
 \item \textbf{H1:} With RGT all positively evaluated aesthetics can be reproduced.
 \item \textbf{H2:} The results of RGT can be reproduced when using different elements and different participants. 
\end{itemize}

\textbf{H1} is used to check whether the RGT provides valid results.
With \textbf{H2}, we check whether the results are reproducible or depend on the selected elements or participants.
We are not aware of any other approach to systematically explore the design space of aesthetics.
A comparative evaluation with other approaches is therefore not possible.

\subsection{Ground truth}

To verify the results of our evaluation, we have conducted an extensive literature study following the guideline from vom~Brocke~et al.~\cite{VomBrocke2009} to establish a ground truth for \textbf{H1}.
It contains all the aesthetics proposed in the literature and whether a significant influence on readability could be empirically evaluated.
We have searched the databases available to us with the search terms listed in Table~\ref{tab:searchCriteria}.

\begin{table*}[tb]
\label{tbl:criteria}
\caption{In- and exclusion criteria for literature study}\label{tab:searchCriteria} \centering
\begin{tabular}{c|c|c}
  Database & Search Term & Inclusion (+) and Exclusion (–) Criteria \\
  \hline
  ScienceDirect & graph aesthetics & \makecell[cl]{+ Publication Type: \textit{Research Article} \\
                                     + Journal: \textit{Computer Aided Design} \\
                                     + Journal: \textit{Journal of Visual Languages}} \\
  \hline
  ACM & (+graph +aesthetics) & \\
  \hline
  IEEE & graph aesthetics & \makecell[cl]{– Publication Type: \textit{Book}} \\
  \hline
  SpringerLink & graph aesthetics & \makecell[cl]{+ Publication Type: Conference Papers \\ 
                                    + Discipline: \textit{Computer Science} \\ 
                                    + Subdiscipline: \textit{Information Systems Appl.} \\ 
                                    + Subdiscipline: \textit{User Interfaces and HCI}} \\
\end{tabular}
\end{table*}

The additional inclusion and exclusion criteria are necessary because the term aesthetics is used in many different disciplines with different meanings.
% Without these criteria we would get thousands of false positive results.
In total, we received 519 hits, 47 from ScienceDirect, 69 from ACM, 42 from IEEE, and 373 from SpringerLink.
Two entries had to be removed due to duplicates, leaving 517 entries.
We then manually sorted out the publications where the term \emph{aesthetics} is not used in the sense mentioned here.
Then, we performed a backward search on the 95 remaining publications. 
This was necessary because many publications use aesthetics, but it was not the original source in which the metric was proposed.
We also included the summaries from Taylor~\cite{Taylor2005} and Bennett~\cite{Bennett2007}, who did a similar literature study with a smaller focus.
The first three columns of Table~\ref{tab:overview} summarize the results of our literature study.
All in all, we identified 29 different graph aesthetics proposed in 14 different publications.
For 13 aesthetics we could find an empirical evaluation that showed a significant influence on readability.
For some aesthetics, we were not able to trace them back to exactly one source.
% Some authors only refer to ``generally accepted'' aesthetics without further reference.
In such a case we listed all found publications.
% This seems to support our assumption, that the desgin space for aesthetics is not researched systematically.

Most aesthetics refer to the position of the nodes, edge intersections, the length and curvature of the edges, and the angles between them.
Some aesthetics refer to paths, i.e., combinations of edges.
For example, path bendiness describes how straight a path is or how many bends it has.
Most aesthetics can be sorted into the ``Composition'' category since they refer to layouting.
Only a few aesthetics belong to the ``Visual Mapping'' category, they are highlighted in Table~\ref{tab:overview}.

% \definecolor{VisualMapping}{rgb}{0.96,0.80,0.38} % orange
\definecolor{VisualMapping}{rgb}{0.63,0.84,1} % hell blau

\begin{table*}[bt]
\caption{List of all aesthetics derived from literature. 
Entries of the category ``Visual Mapping'' are highlighted.}\label{tab:overview} \centering
\begin{tabular}{l|l|l|c|c|c}
  Name & Source & Evaluation & Group A & Group B & Group C \\
  \hline
%   \rowcolor{Composition}
    Angular resolution & \cite{Purchase2002,Taylor2005,Battista1999} & \cite{Huang2010} & 4 & 3 & 3 \\
% % %   \rowcolor{Composition}
    Area & \cite{Tamassia1988,Taylor2005} & \cite{Purchase2002a} & 10 & 8 & 8 \\
%   \rowcolor{Composition}
    Aspect ratio & \cite{Battista1999} & & 3 & 4 & 3 \\
%   \rowcolor{Composition}
    Cluster similar nodes & \cite{Tamassia1988,Taylor2005} & \cite{Huang2014} & 5 & 5 & 4 \\
%   \rowcolor{Composition}
    Convex faces & \cite{Tamassia1988} & & – & – & – \\
%   \rowcolor{Composition}
    Consistent flow direction & \cite{Purchase2002} & & 3 & 4 & 6 \\
%   \rowcolor{Composition}
    Crossing angle & \cite{369,Ware2002,Huang2010} & \cite{Ware2002,Huang2010} & 8 & 9 & 7 \\
  \rowcolor{VisualMapping}
    Degree of edge bends & \cite{Purchase2002,Tamassia1988,DH96} & \cite{Purchase1997,Purchase1997a,Purchase2002a} & 9 & 9 & 10 \\
%   \rowcolor{Composition}
    Difference between angles & \cite{Hutchison} & & – & – & – \\
%   \rowcolor{Composition}
    Distribute nodes evenely & \cite{Tamassia1988,Taylor2005} & & 6 & 8 & 8 \\
  \rowcolor{VisualMapping}
    Edge orthogonality & \cite{Purchase2002} & \cite{Purchase2002a} & 5 & 4 & 4 \\
%   \rowcolor{Composition}
    Global symmetry & \cite{Tamassia1988,Biedl1999} & \cite{Purchase1997} & 4 & 3 & 4 \\
%   \rowcolor{Composition}
    Keep nodes apart from edges & \cite{DH96} & & 3 & 6 & 7 \\
%   \rowcolor{Composition}
    Local symmetry & \cite{Tamassia1988,Biedl1999} & \cite{Purchase2002a} & 8 & 10 & 8 \\
%   \rowcolor{Composition}
    Maximum bends & \cite{Battista1999} & & 9 & 9 & 8 \\
%   \rowcolor{Composition}
    Maximum edge length & \cite{Battista1999,Taylor2005,Tamassia1988} & & 6 & 4 & 4 \\
%   \rowcolor{Composition}
    Node orthogonality & \cite{Purchase2002} &  & – & 3 & – \\
%   \rowcolor{Composition}
    Nodes should not overlap & \cite{Wetherell1979} & & 4 & 3 & 3 \\
%   \rowcolor{Composition}
    Number of bends & \cite{Battista1999} & & 3 & 3 & 4 \\
%   \rowcolor{Composition}
    Number of branches & \cite{Ware2002} & \cite{Ware2002} & 5 & 3 & 5 \\
%   \rowcolor{Composition}
    Number of edge crossings & \cite{Tamassia1988,Taylor2005,Purchase2002,DH96,Biedl1999} & \cite{Purchase1997,Purchase1997a,Purchase2002a,Polisciuc} & 6 & 3 & 8 \\
%   \rowcolor{Composition}
    Path bendiness & \cite{Ware2002}  & \cite{Ware2002} & 3 & 3 & 5 \\
%   \rowcolor{Composition}
    Shortest path length & \cite{Ware2002} & \cite{Ware2002} & 4 & 3 & 3 \\
%   \rowcolor{Composition}
    SD of crossing angles & \cite{Huang2010} & & – & – & – \\
%   \rowcolor{Composition}
    SD of angular resolution & \cite{Huang2010} & & – & – & – \\
%   \rowcolor{Composition}
    Total edge length & \cite{Tamassia1988,Taylor2005} & & – & – & –\\
  \rowcolor{VisualMapping}
    Uniform edge bends & \cite{Taylor2005} & & 3 & 3 & 4 \\
  \rowcolor{VisualMapping}
    Uniform edge lengths & \cite{Tamassia1988,Gansner,Biedl1999,DH96} & & 4 & 3 & 3 \\
%   \rowcolor{Composition}
    Whitespace to ink ratio & \cite{Polisciuc,Tullis2016} & \cite{Polisciuc} & 3 & 3 & 6 \\
\end{tabular}
\end{table*}

\subsection{Study Design} \label{sec:StudyDesign}

\subsubsection{Elements}

For each group, we used 12 undirected graphs as elements.
They can be seen in the appendix.
% They can be found in the full version [99].
They consist only of black nodes and black undirected edges.
We did not use any text labels or color mappings to keep the graphs as simple as possible.
The graphs are not based on real but on artificially generated data.
We used the \textit{igraph} library for R\footnote{https://igraph.org/r/} to generate random graphs.
The smallest graph contains 5 edges, the largest graph contains 69 edges.
Each node position was assigned randomly, i.e., overlaps could and did occur.
For each edge, the degree and direction of edge curvature were determined randomly as well as which nodes the edge connects.
No other properties were taken into account.
Fig.~\ref{fig:grid_ui} shows three of the used graphs.

\subsubsection{Participants}

In total, we interviewed 30 participants.
Initially, these participants were divided into three groups to check \textbf{H2}.
We decided on a group size of 10 because it has proven to be sufficient in many studies.
If the method is widely applied, it may be possible to find a convergence point at which additional participants do not add any value.
All participants were bachelor or master students of economics and have received an expense allowance.
They were all native speakers of German, which was also the language of the interviews.
In group A, the students were between 19 and 40 years old (mean: 23.3 years). 
50\% were female, 50\% male.
In group B, the students were between 18 and 29 years old (mean: 21.9 years).
40\% were female, 60\% male.
In group C, the students were between 19 and 25 years old (mean: 21.5 years).
60\% were female, 40\% male.
Participants of the same group have worked with the same elements.

\subsubsection{Interview}

\begin{figure}[!bt]
  \centering
  \includegraphics[width=\textwidth]{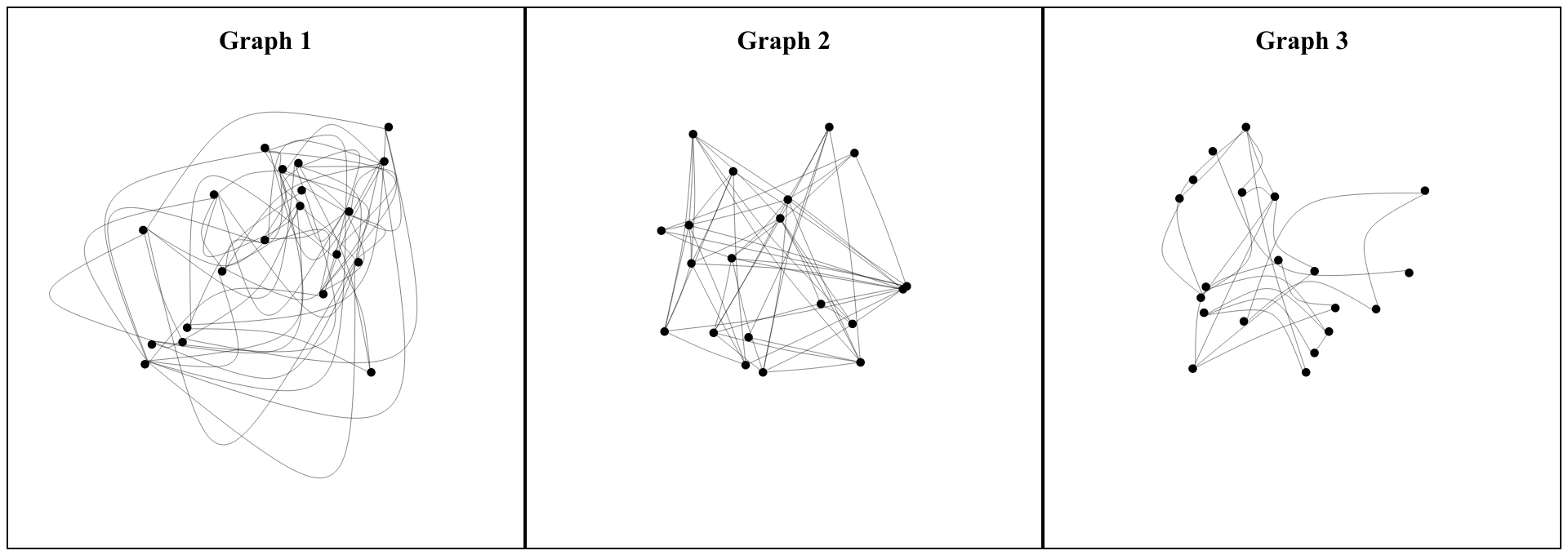}
  \caption{User Interface for Repertory Grid Interview showing three graphs}
\label{fig:grid_ui} 
\end{figure}

The complete evaluation was done using the evaluation server of Getaviz~\cite{Baum2017}.
It displays three random graphs at the same time (see Fig.~\ref{fig:grid_ui}).
The participant cannot interact with the visualizations, i.e., there are no tooltips and it is not possible to navigate or zoom in and out.
In the prestudy, we noticed that sometimes rather vague terms such as ``simple'' or ``complex'' were used as constructs.
To improve the laddering, we asked the participants to draw for instance a ``very simple'' or ``very complex'' graph and used it as an additional element.
Having additional elements with extreme properties helps the participant to name differences between the elements~\cite{Kelly1991}.
Besides that, we conducted the interview as described in the method section.

\subsection{Results and Discussion}

\begin{table}[tb]
\caption{List of all novel aesthetics elicited in the evaluation}\label{tab:novel} \centering
\begin{tabular}{l|c|c|c}
  Name & Group A & Group B & Group C \\
  \hline
  Face area     & 2 & 3 & 3 \\
  Uniform faces & 4 & 3 & 3 \\
\end{tabular}
\end{table}

The interview procedure led to a set of 56 different constructs from all participants. 
% Each participant used 25.3 constructs on average, without significant differences between the three groups.
These constructs are divided into the four categories as follows:
Visual Mapping (4 constructs), Composition (21 constructs), Data-related (11 constructs), Visual Experience (20 constructs).
The distribution of the categories is similar to previous studies but with fewer constructs referring to visual mapping~\cite{Kurzhals2018}.
That was to be expected since the visual mapping was given by using node-link diagrams and corresponds to the distribution of published graph aesthetics, which refer to the composition in most cases as well.
The further analysis will focus on the 25 constructs from the first two categories since the other constructs are not relevant concerning aesthetics.
For each aesthetic in Table~\ref{tab:overview} it is indicated which groups have used it.
We can fully confirm hypothesis \textbf{H1}.
An aesthetic was used by 51.7\% of the participants on average (min: 33.3\%, max: 93.3\%)
With a softer stop criterion, some aesthetics might have been used by more participants.
It is neither necessary nor likely that all participants use identical constructs.

We were able to reproduce all published graph aesthetics that have an empirically verified impact on readability with all three groups.
Group~A reproduced 82\%, Group~B reproduced 86\%, and Group~C reproduced 82\% of published graph aesthetics.
The five aesthetics not mentioned were not positively evaluated without exception.
In the case of \textit{differences between smallest and optimal crossing angle}, \textit{standard deviation of crossing angles}, and \textit{standard deviation of angular resolution} this is not surprising.
Participants of all groups referred to crossing angles quite often, but not in such a mathematical way.

\begin{figure}[!b]
  \centering
   \subfloat[``Ideal Graph'']{
    \includegraphics[width=2in]{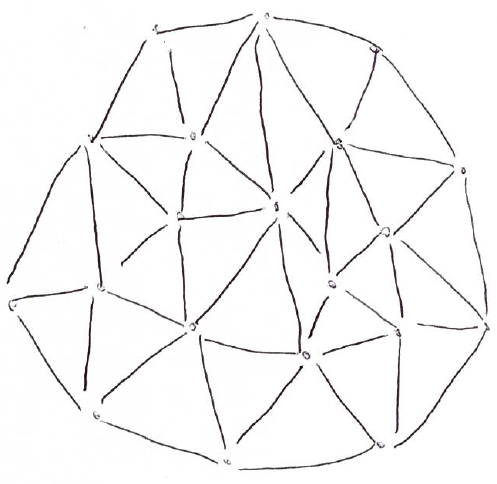}
    \label{fig:ideal_graph}
  }
  \hfill
  \subfloat[``Worst Graph'']{
    \includegraphics[width=2in]{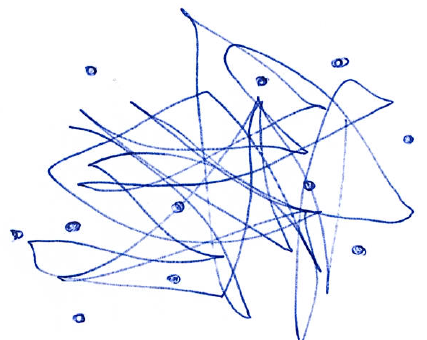}
    \label{fig:worst_graph}
  }
\caption{Two example graphs drawn by participants}
\label{fig:drawings}
\end{figure}

Table~\ref{tab:novel} lists all elicited aesthetics that are novel, which means that we could not find a corresponding aesthetic in our literature study.
Both novel aesthetics refer to faces, i.e., the empty white areas that are bordered by edges.
So far in the literature, it has only been suggested to consider whether the faces are convex or concave.
This was not relevant for any of the participants.
However, participants of all groups distinguished between faces with a small area and faces with a huge area.
They also took into account, whether the graph consists of faces with a similar shape or not.
The results of our evaluation indicate that the area and shape of the faces might influence how a graph is perceived.
It has to be verified empirically whether these aesthetics have a significant impact on understandability and readability.

% The confirmation of \textbf{H1} implies, that all three groups produced identical results with respect to already evaluated aesthetics.
With one exception, the used aesthetics are consistent among all three groups.
Only participants of Group~2 used node orthogonality to differentiate between the elements.
% Looking at the other aesthetics, small differences are apparent.
% Only for two aesthetics the groups produced different results: \textit{path bendiness} and \textit{shortest path length}
% Group~1 proudces two aesthetics, that were not mentioned by subjects of Group 2 and 3.
% It has to be checked empirically if these aesthetics are relevant.
% We can not say with the existing data whether it is false positive or false negative.
% However, it is noticeable that the differences relate to the only two aesthetics that refer to paths.
% Paths were not taken into account in the random generation of the graphs.
% Group~2 produces also one unique aesthetic.
% Other than that, the results are consistent.
% 89\% of the elicited aesthetics are identical.
Therefore, we can accept hypothesis \textbf{H2} conditionally.

\subsection{Threats to Validity}

For the interviews, we have specified the elements and deliberately used random values for different properties of the graphs.
There is a risk that thereby the aesthetics are predetermined and reflect only our assumptions.
However, the participants mentioned aesthetics that have no direct connection to the randomized graph properties.
For example, all groups used \emph{global symmetry} as a construct.
None of the given elements was symmetric or designed with respect to symmetry.
However, many self-drawn graphs were symmetrical as shown in~\ref{fig:ideal_graph}, making them different from the elements provided.

All interviews were conducted by the same person, therefore there is a risk of confirmation bias.
Other potential confounding factors are the background and degree of experience of the participants.

\section{Application to Software Visualization}

\begin{figure}[!b]
    \centering
    \subfloat[RD Metaphor]{
        \includegraphics[width=2in]{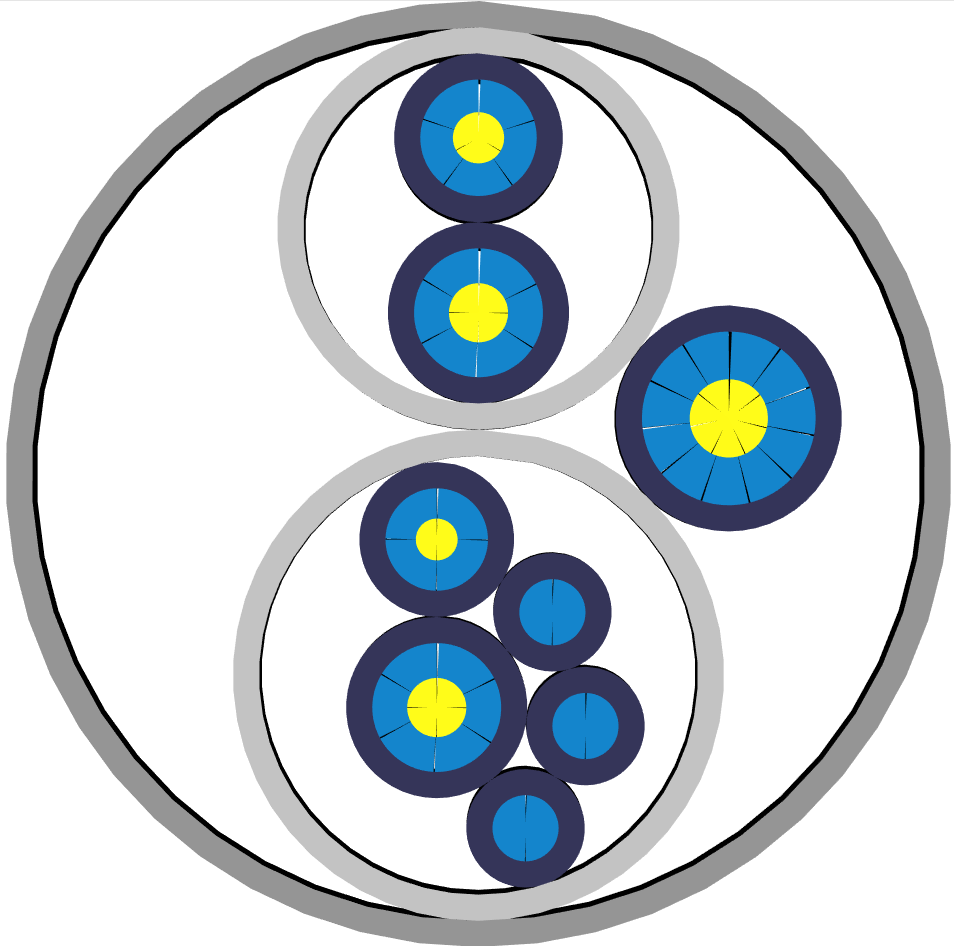}
        \label{fig:RD}
    }
    \hfill
    \subfloat[City Metaphor]{
        \includegraphics[width=2in]{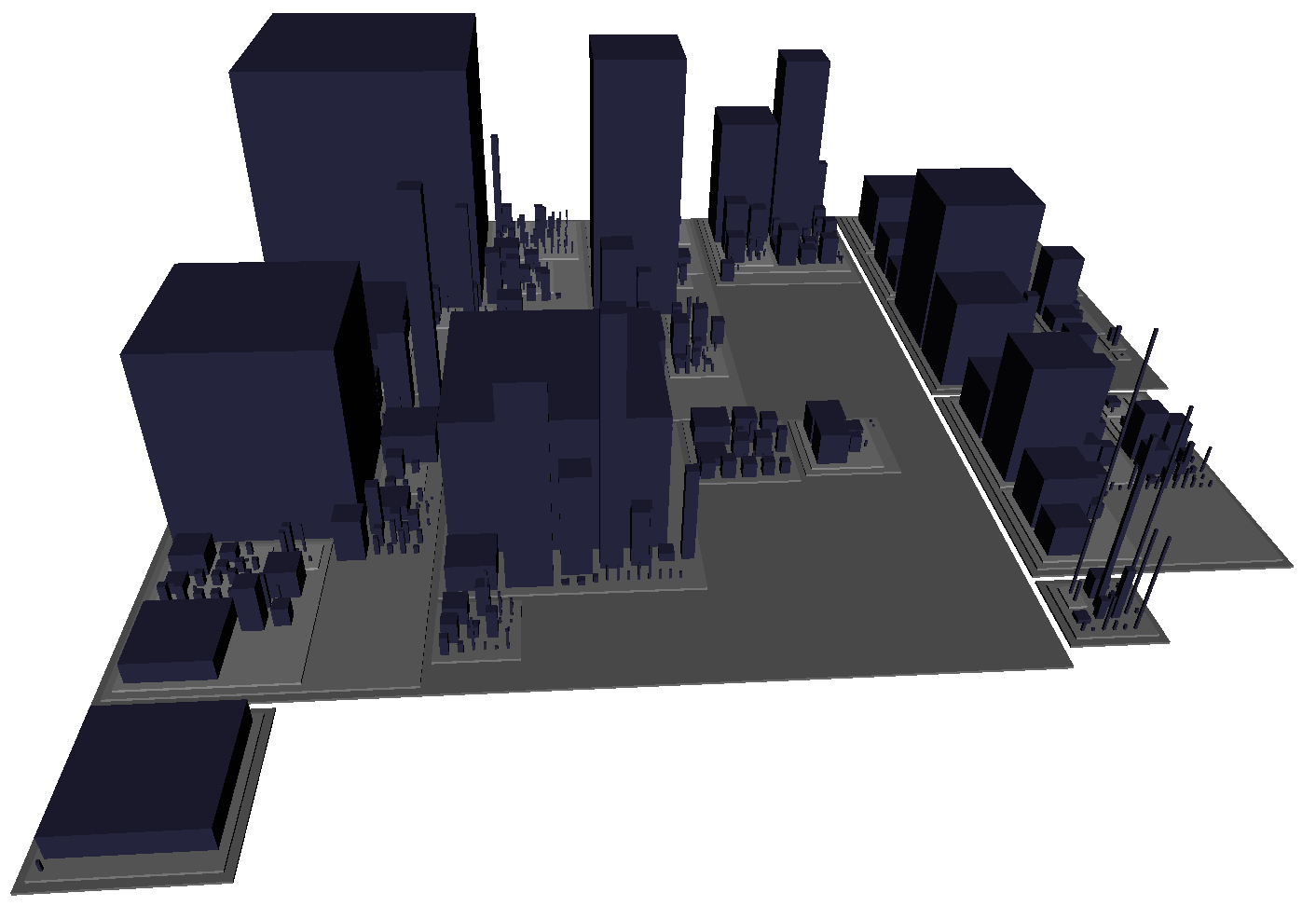}
        \label{fig:City}
    }
    \caption{Software visualizations generated by Getaviz}
    \label{fig:metaphors}
\end{figure}

Software visualization is a subdomain of information visualization about visualizing the structure, behavior, and evolution of software systems.
These visualizations are used in visual analytics tools to support software developers, project managers, and other stakeholders to improve their understanding of development artifacts and corresponding activities.
Software visualizations are complex domain-specific diagrams that might contain multiple thousand data points, various relationships between them, and a multitude of different visual primitives.
Presenting this amount of information in such a way that it can be processed well by a human being is a central challenge of this domain.
The Recursive Disk (RD) Metaphor (Fig.~\ref{fig:RD})~\cite{Mueller2015b} and the City Metaphor (Fig.~\ref{fig:City})~\cite{Wettel2008c} are two approaches to adequately visualize these data.

Both metaphors are hierarchical visualizations that represent the internal structure of a software system.
The RD Metaphor is an abstract two-dimensional metaphor.
It consists of two different kinds of disks (gray and purple) as well as two different kinds of disk segments (blue and yellow).
The disks can be nested to represent contains-relationships between the elements as shown in Fig.~\ref{fig:RD}.
The area of the disks and disk segments is also used to visualize the properties of the software system.
The City metaphor is a three-dimensional real-world metaphor. 
It consists of gray districts and purple buildings as shown in Fig.~\ref{fig:City}.
The building's height and base area also represent the properties of the software system.

A high degree of readability and comprehensibility is a central requirement for these kinds of diagrams.
To improve them, however, no aesthetics have been considered to date, i.e., there are no known aesthetics at all for this kind of visualization.
One of the reasons for this is that the described problems of the current research process are even greater with such complex visualizations.
Therefore, we apply our approach to software visualizations to elicit aesthetics that will help improve readability and comprehensibility in the future.
We have conducted one study on RD visualizations and one on City visualizations.
Both studies are independent of each other.
However, since the study design is very similar, we will describe both studies together.

\subsection{Study Design}

For each study, we used 12 visualizations as elements.
We chose 12 different software systems based on software metrics (number of packages, number of classes, number of methods, number of attributes, and number of statements) to cover a wide range.
We used Getaviz to generate the corresponding visualizations for each system. 
For RD, we conducted interviews with ten participants (50\% male, 50\% female).
Their age varies between 19 and 52 years (mean: 22.8 years).
For City, we conducted interviews with ten different participants (70\% male, 30\% female).
Their age varies between 18 and 38 years (mean: 24 years).
During the construct elicitation we gave participants the possibility to navigate, i.e., rotate the visualization as well as zoom in and out, so they could view the visualization as they liked.
Otherwise, it would not be possible to perceive all visual entities since some entities might be occluded or too small to perceive.
Apart from that, the interviews were the same as described in Section~\ref{sec:StudyDesign}.

\subsection{Results and Discussion}

We elicited 53 constructs during the RD interviews, 19 of them qualified as aesthetics.
Each participant used 15.5 constructs on average.
To describe the city metaphor, 45 constructs were used, 15 of them are aesthetics.
Each participant used 13.5 constructs on average.
Table~\ref{tab:softvis_aesthetics} lists all elicited aesthetics for both metaphors.
The aesthetics for the RD visualizations are mostly about color distribution and nesting, i.e., many aesthetics refer to a local context.
This makes sense considering the recursive structure of the visualizations.
% Subjects used the terms \textit{strings} to describe certain visual patterns.
% Fig.~\ref{fig:terms} illustrates what is meant by these terms.
City visualizations have a similar structure, but buildings are clearly dominant since most aesthetics refer to buildings.
Some aesthetics refer to the three-dimensionality of the visualization, where the height of the buildings plays a major role.

\begin{table*}[tb]
\caption{Elicited Aesthetics for RD and City Metaphor}\label{tab:softvis_aesthetics} \centering
\begin{tabular}{l|l}
  \makecell[c]{RD Aesthetics} & \makecell[c]{City Aesthetics} \\
  \hline
    Area & Area \\
    Blue segments evenly distributed (global) & Aspect ratio (global) \\
    Blue segments evenly distributed (local) & Aspect ratio of districts (local) \\
    Centered focus & Buildings in a row \\
    Edge thickness & Building density \\
    Face area & Clustering of similar buildings \\
    Global symmetry & Empty district area \\
    Length of spiral windings & Gap between buildings \\
    Local symmetry & Largest difference in building height \\
    Nesting depth & Nesting depth \\
    Number of spiral turns & Share of empty area \\
    Share of empty area & Sort buildings by height \\
    Sorting of purple disks (local) & Uniform base area of buildings \\
    Uniform size of gray disks & Uniform buildings \\
    Uniform size of purple disks & Uniform faces \\
    Uniform structure of gray disks & \\
    Uniform structure of purple disks & \\
    Yellow segments evenly distributed (global) & \\
    Yellow segments evenly distributed (local) & \\
\end{tabular}
\end{table*}

It is particularly noticeable that fewer constructs were used compared to graphs, both per interview and overall.
This is most likely because the edges of the graphs have many degrees of freedom that are not present in the RD and City visualizations.
Due to the semantic constraints, e.g., a building must always be located in a district, the design space is not as extensive as it is for graphs.

The elicited aesthetics serve as a starting point to design better layout algorithms.
In previous work, only density and area were considered.
The elicited aesthetics must now be empirically evaluated to find out which of them have a significant influence on readability.

\section{Conclusion}

Our approach to explore aesthetics design space using repertory grids has been effective.
We have evaluated the approach as far as possible and were able to show in an empirical study that with only 10 participants all published and positively evaluated aesthetics can be identified.
We could also show that our approach delivers reproducible results and can be applied to diverse visualizations.
% However, further studies are necessary to evaluate the correctness of the results for software visualizations.
% In our future work we will evaluate the derived aesthetics from software visualizations to further validate the results.
The quality and validity of the results depend above all on the selection of the suitable elements.
The inclusion of drawings and placeholder elements was particularly helpful.
However, the assessment of a domain expert is still necessary to create and select suitable elements.
Nevertheless, the process is much less subjective and intuition-based than before.

The analysis of the repertory grid data applied in this paper is rather simple and could be enhanced in the future.
For example, we did not analyze how often certain aesthetics have been used by participants.
% This information could be used to determine which aesthetics are particularly important for the observer.
In our future work, we will evaluate the derived aesthetics from software visualizations to further validate the results.
% We hope that our work will help to create visualizations that are easier to read and to understand.

% \section{Conclusion}

% Our approach to explore aesthetics design space using repertory grids has proven to be effective.
% We were able to show in an empirical study that with only 10 subjects all known relevant aesthetics can be identified.
% We could also show that our approach delivers reproducible results and can be applied to diverse visualizations.
% In our future work we will evaluate the derived aesthetics from software visualizations to further validate the results.
% We hope that our work will help to create visualizations that are easier to read and to understand.

\bibliographystyle{splncs04}
\bibliography{literature}

\pagebreak
\section*{Appendix}

\begin{figure}[!ht]
\captionsetup[subfigure]{labelformat=empty}
\centering
\subfloat[Graph 1]{%
  \includegraphics[width=0.3\textwidth,keepaspectratio]{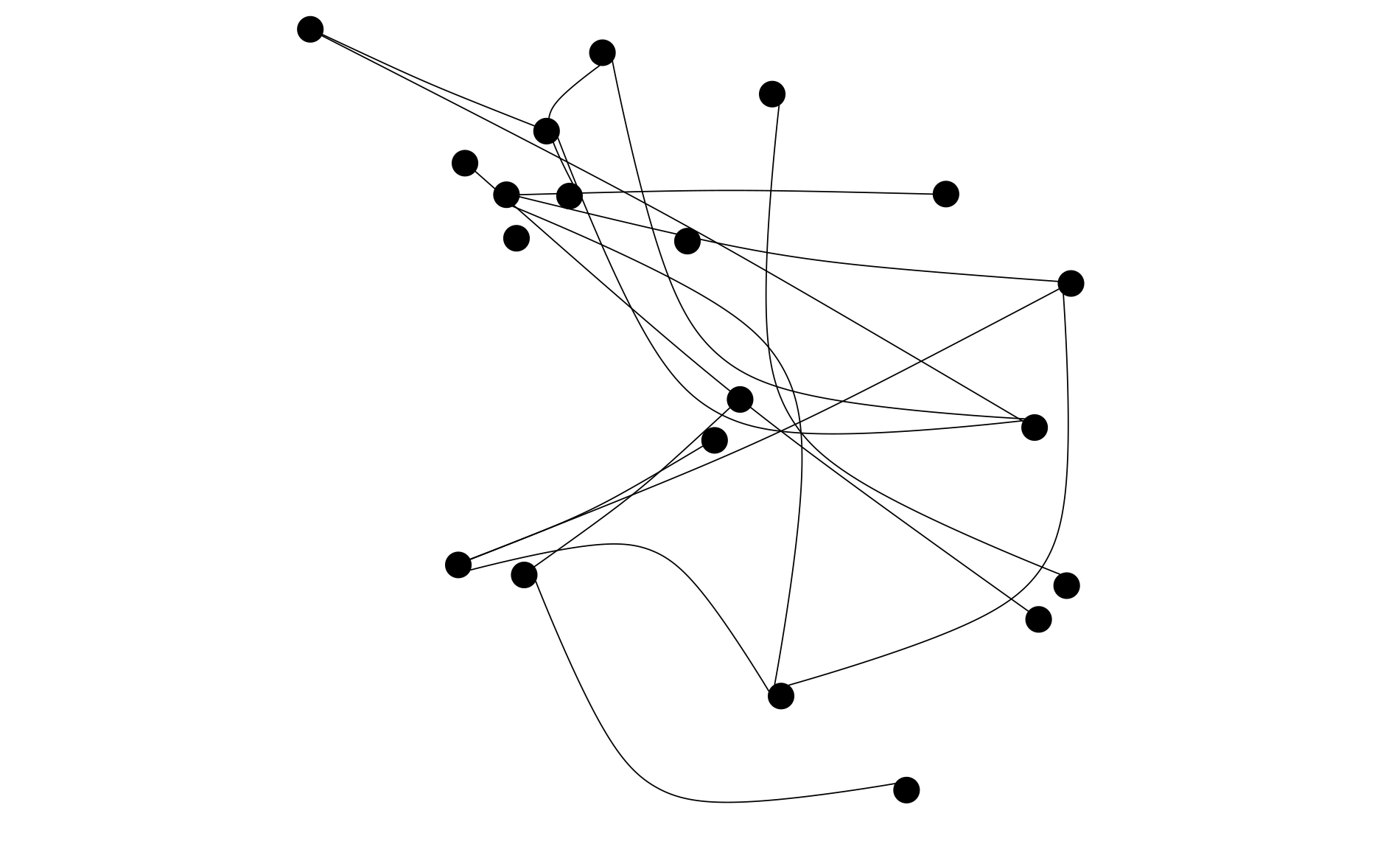}
}\hfill
\subfloat[Graph 2]{%
  \includegraphics[width=0.3\textwidth,keepaspectratio]{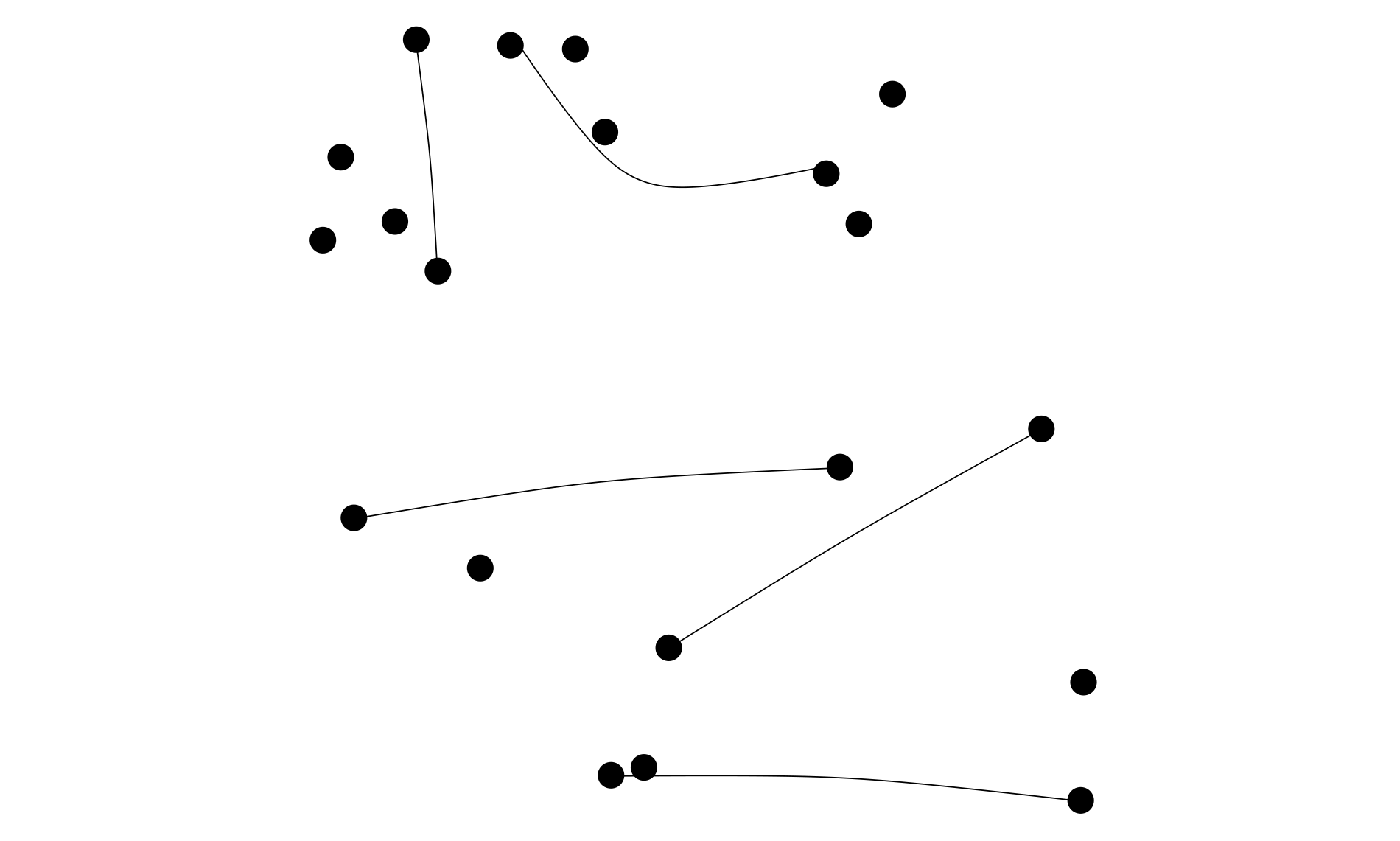}
}\hfill
\subfloat[Graph 3]{%
  \includegraphics[width=0.3\textwidth,keepaspectratio]{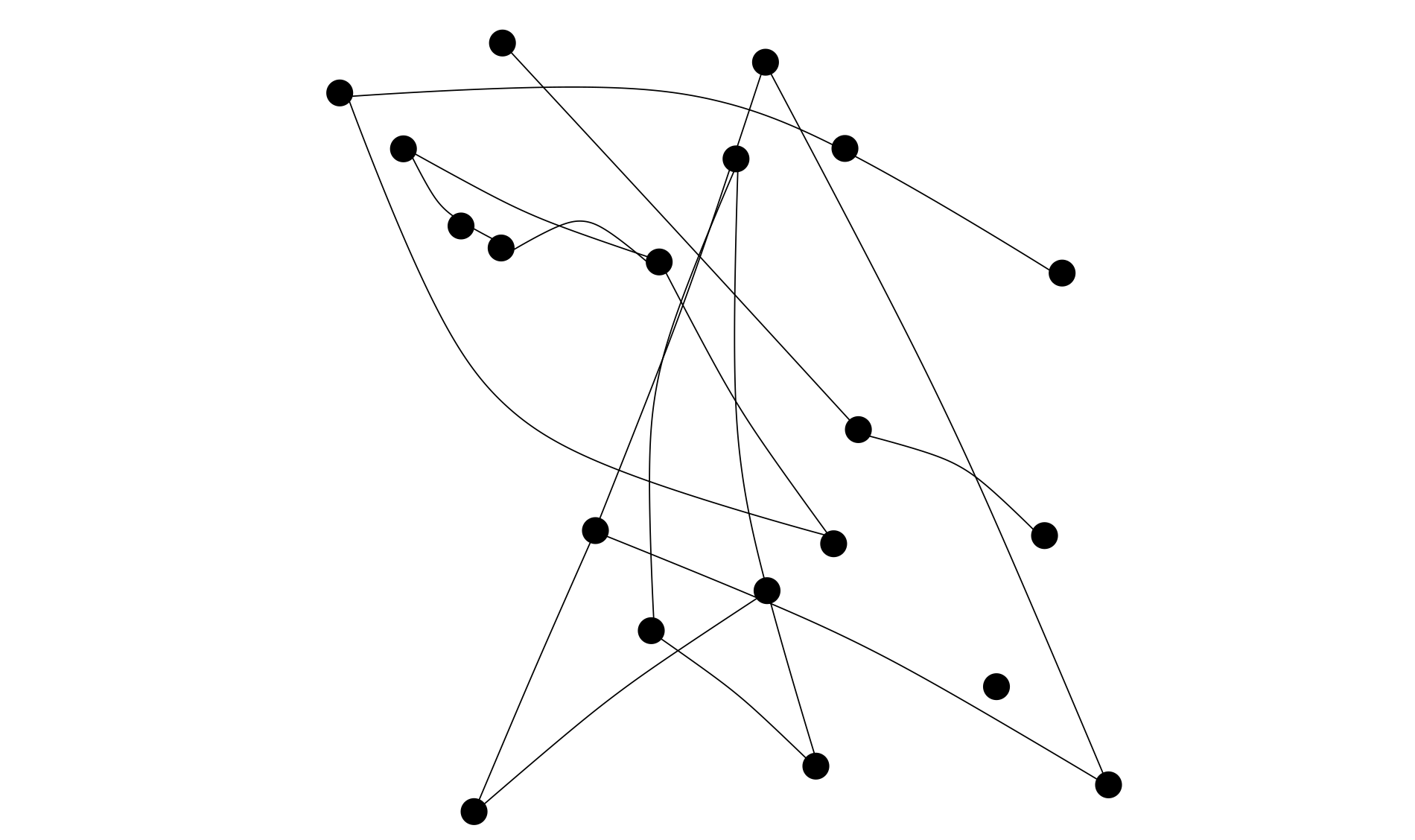}
}\\

\subfloat[Graph 4]{%
  \includegraphics[width=0.3\textwidth,keepaspectratio]{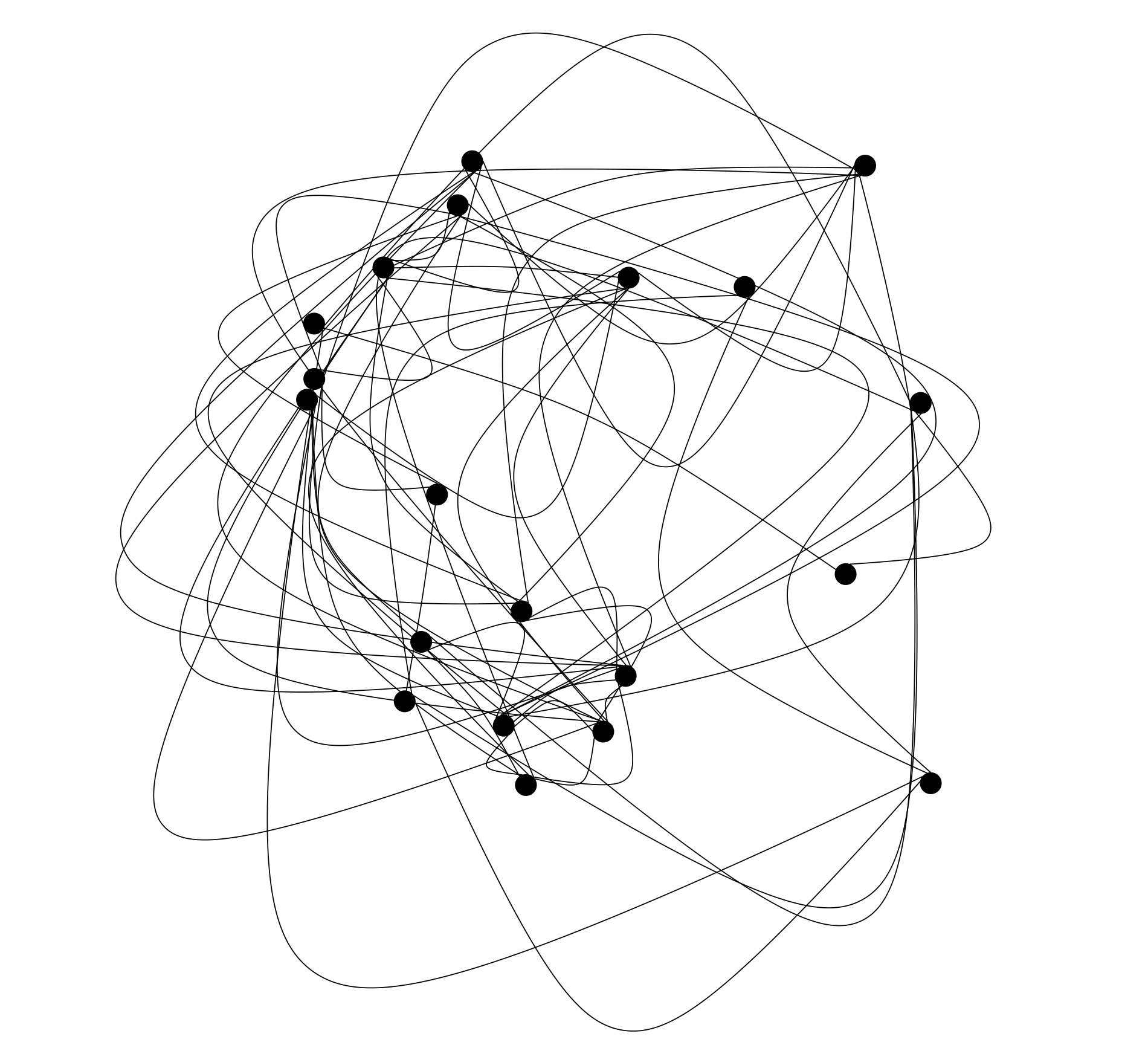}
}\hfill
\subfloat[Graph 5]{%
  \includegraphics[width=0.3\textwidth,keepaspectratio]{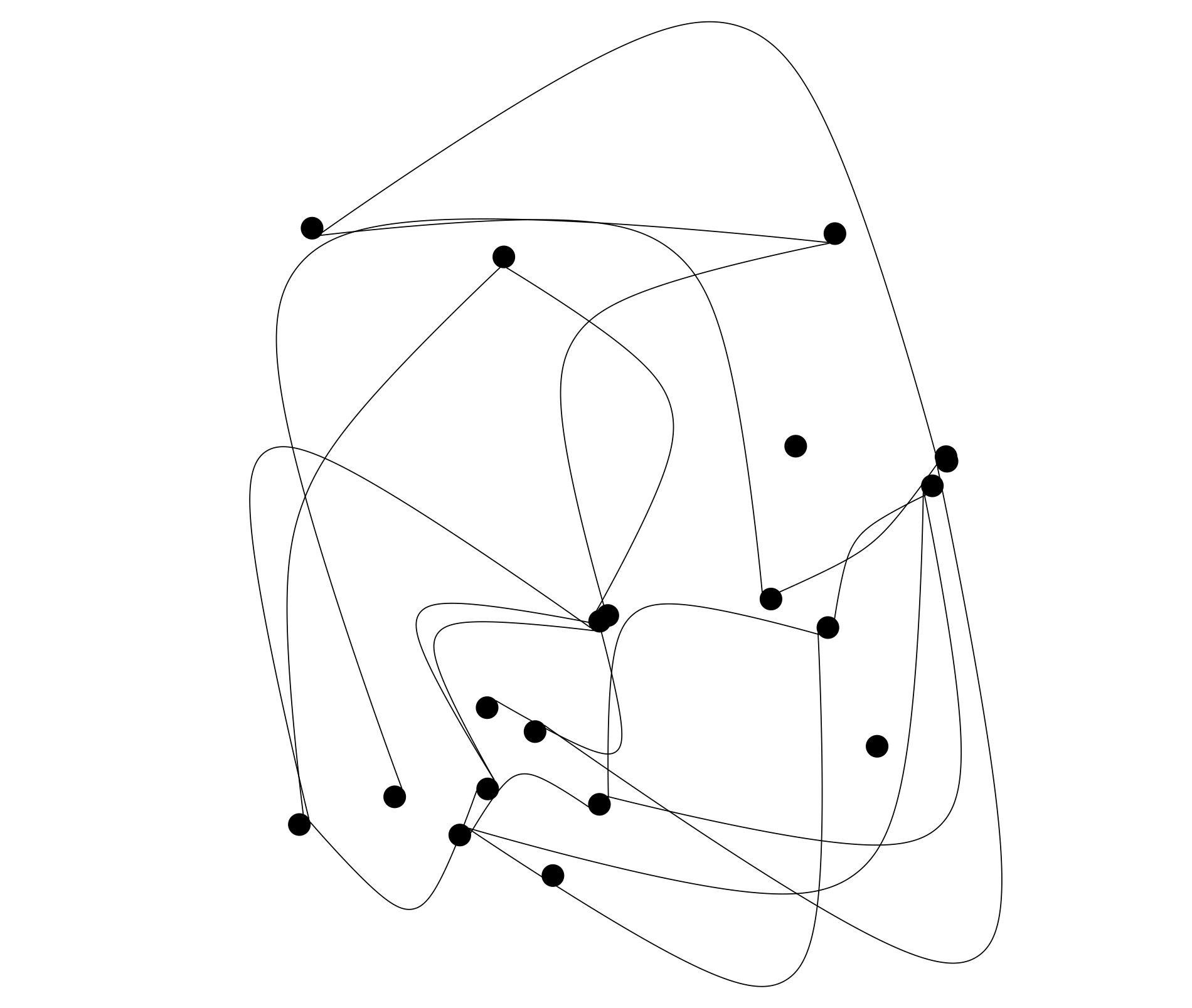}
}\hfill
\subfloat[Graph 6]{%
  \includegraphics[width=0.3\textwidth,keepaspectratio]{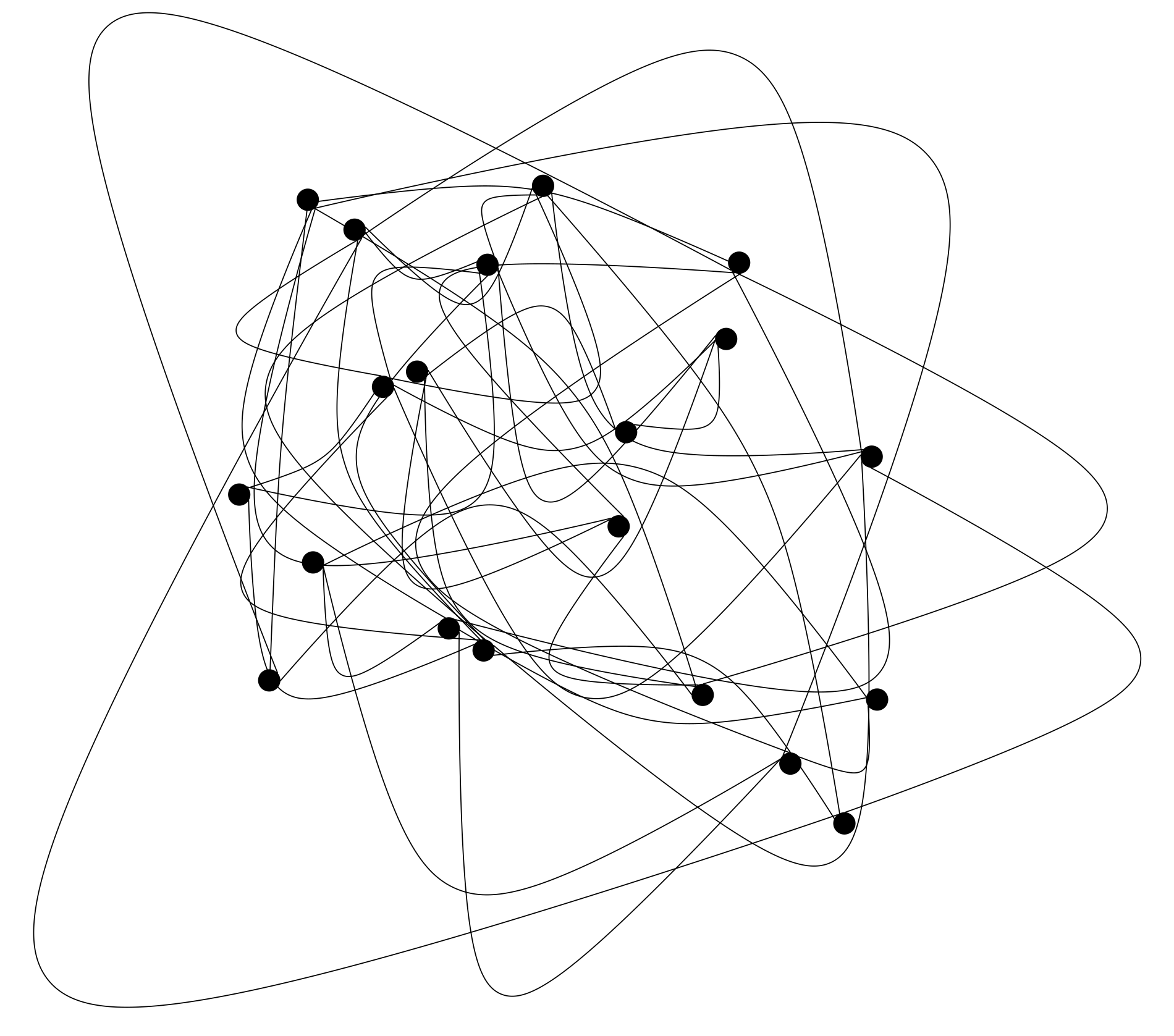}
}\\

\subfloat[Graph 7]{%
  \includegraphics[width=0.3\textwidth,keepaspectratio]{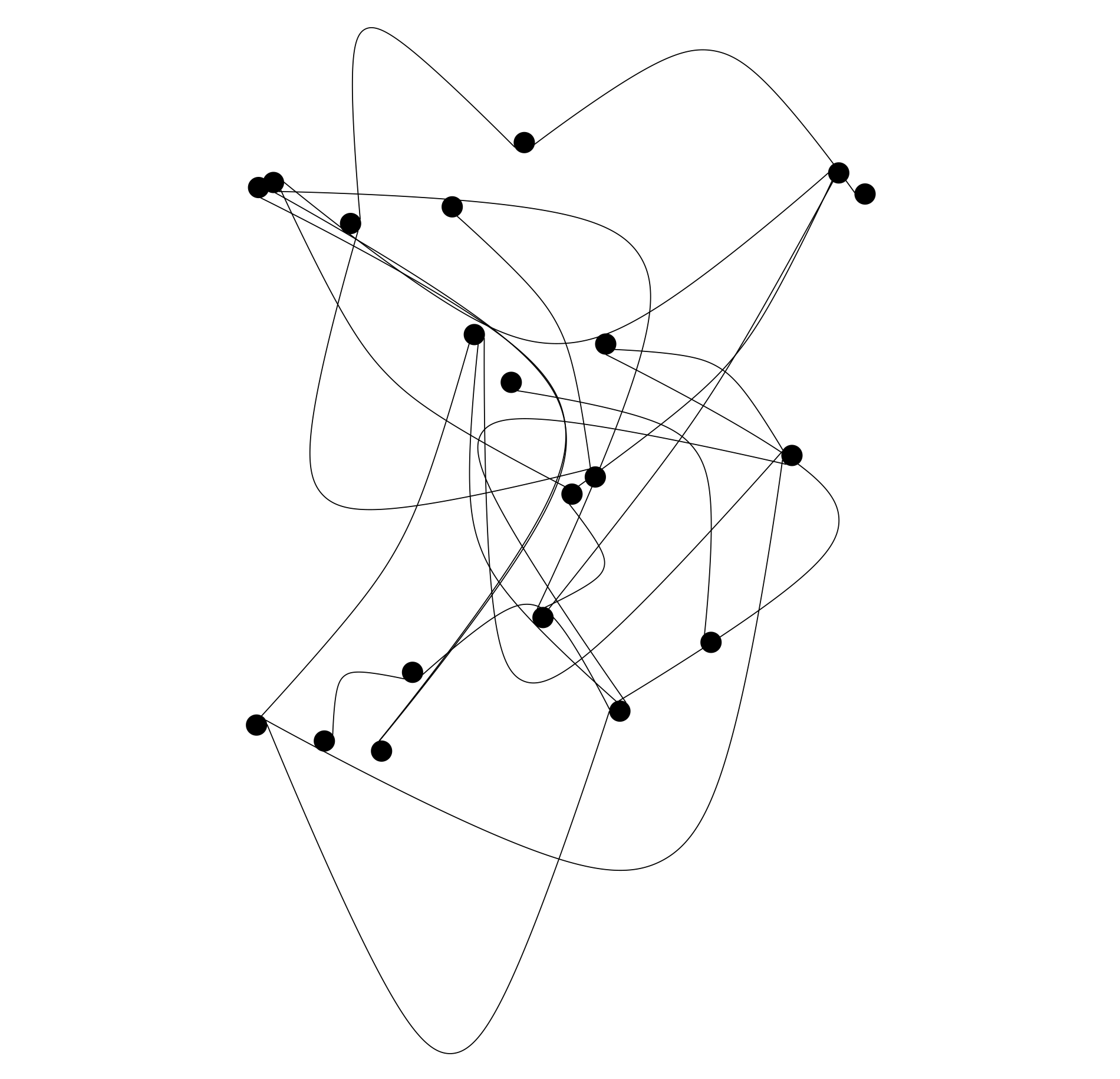}
}\hfill
\subfloat[Graph 8]{%
  \includegraphics[width=0.3\textwidth,keepaspectratio]{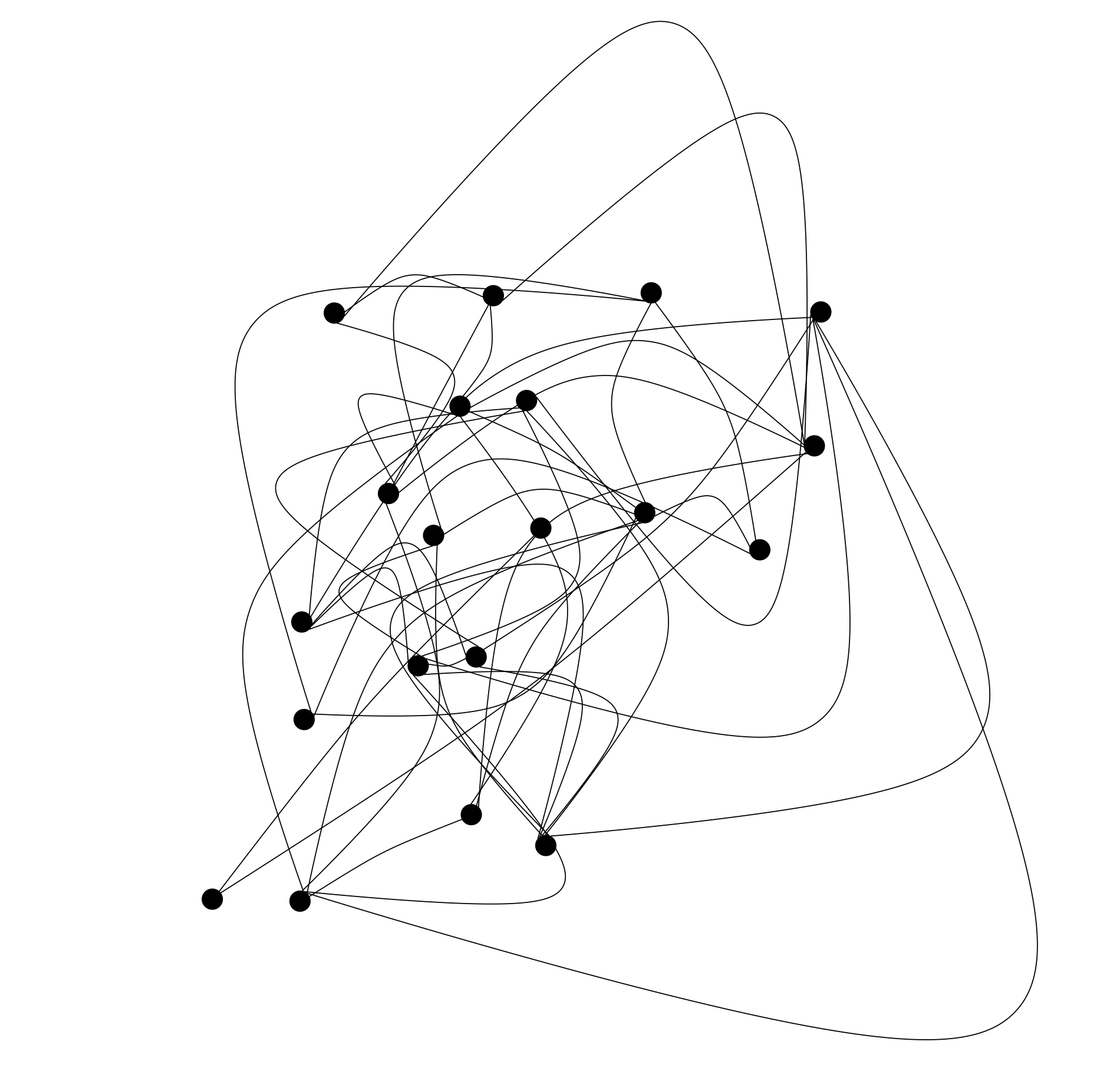}
}\hfill
\subfloat[Graph 9]{%
  \includegraphics[width=0.3\textwidth,keepaspectratio]{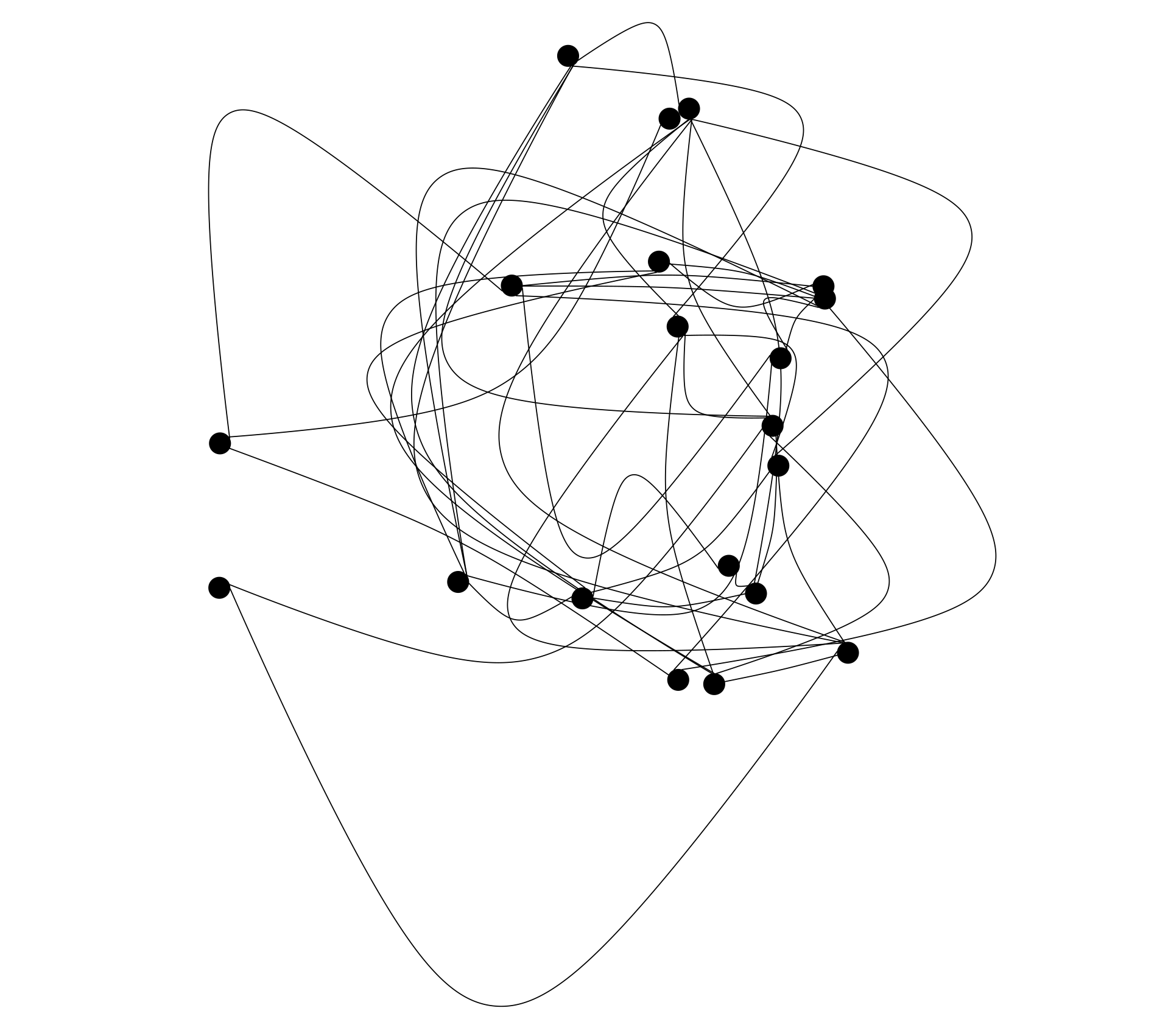}
}\\

\subfloat[Graph 10]{%
  \includegraphics[width=0.3\textwidth,keepaspectratio]{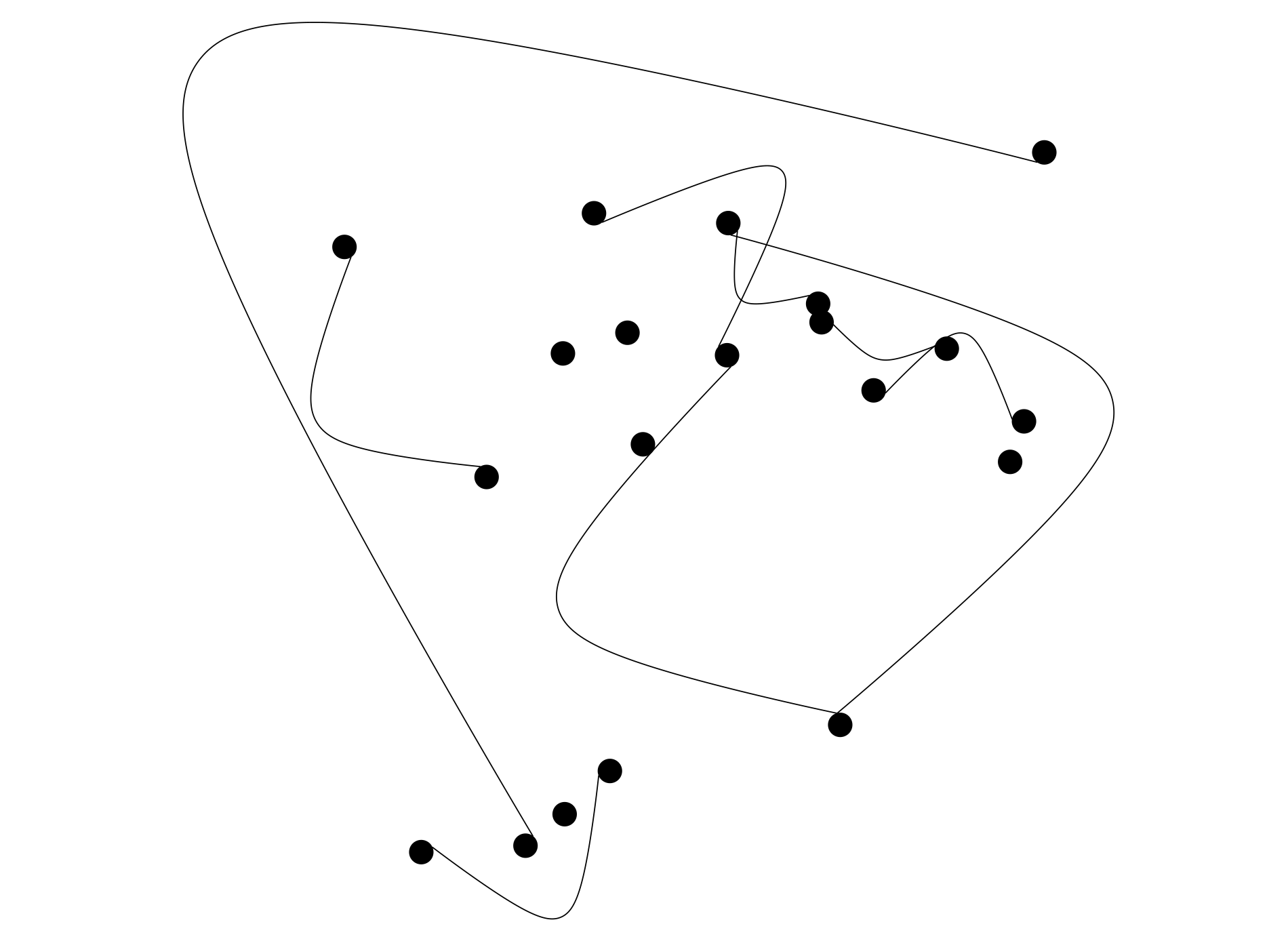}
}\hfill
\subfloat[Graph 11]{%
  \includegraphics[width=0.3\textwidth,keepaspectratio]{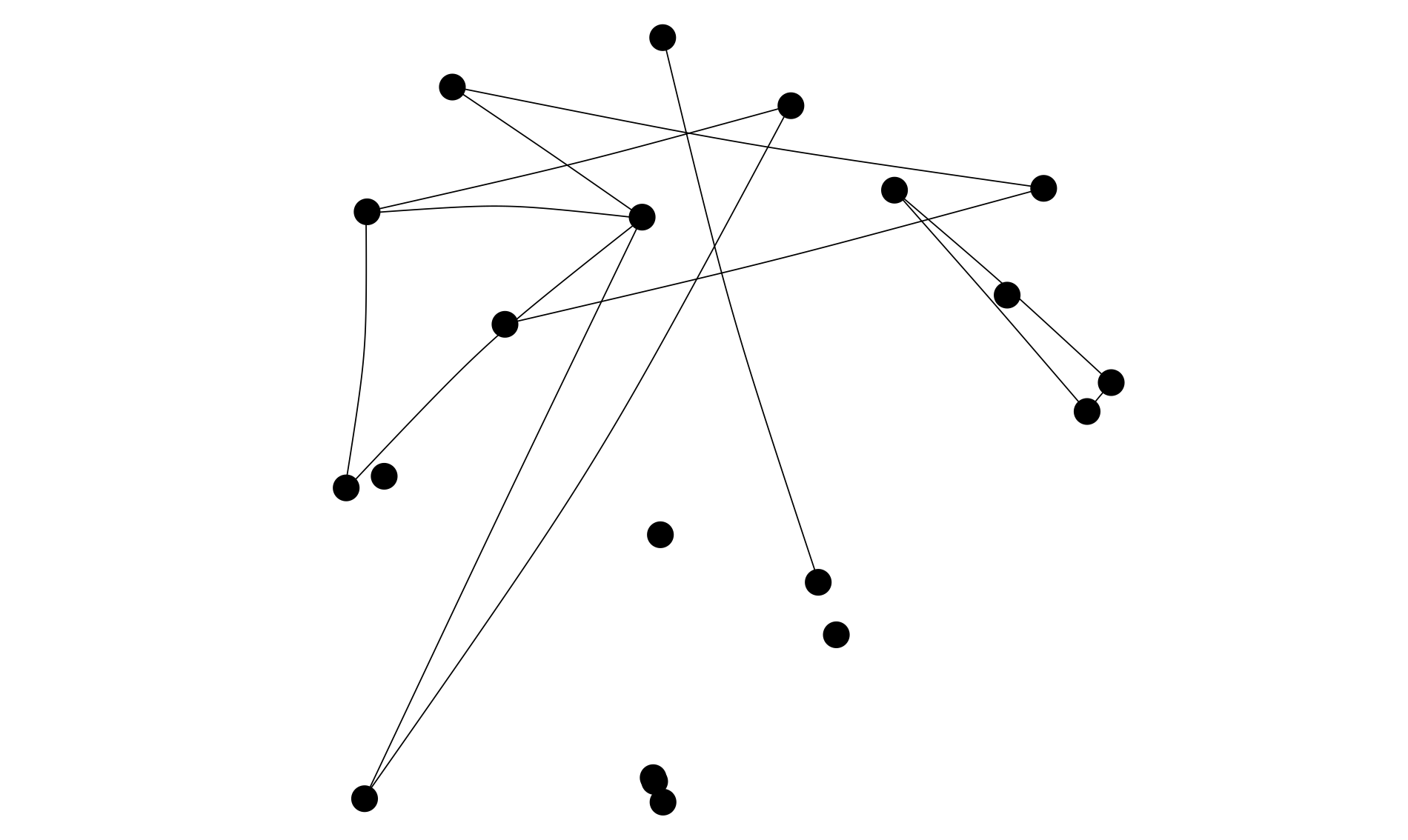}
}\hfill
\subfloat[Graph 12]{%
  \includegraphics[width=0.3\textwidth,keepaspectratio]{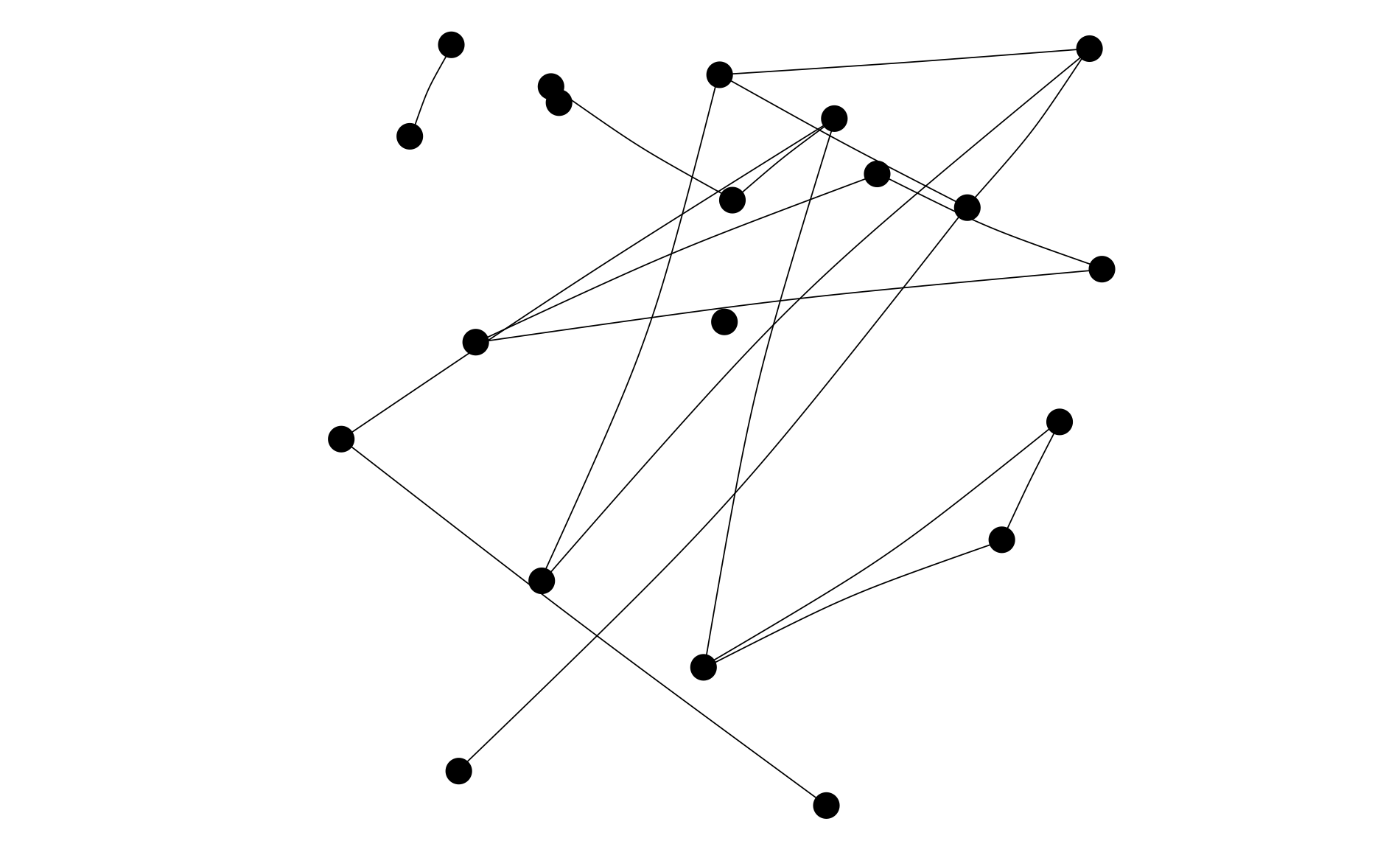}
}
\caption{Elements for Group A}
\end{figure}

\begin{figure}[!th]
\captionsetup[subfigure]{labelformat=empty}
\centering
\subfloat[Graph 1]{%
  \includegraphics[width=0.3\textwidth,keepaspectratio]{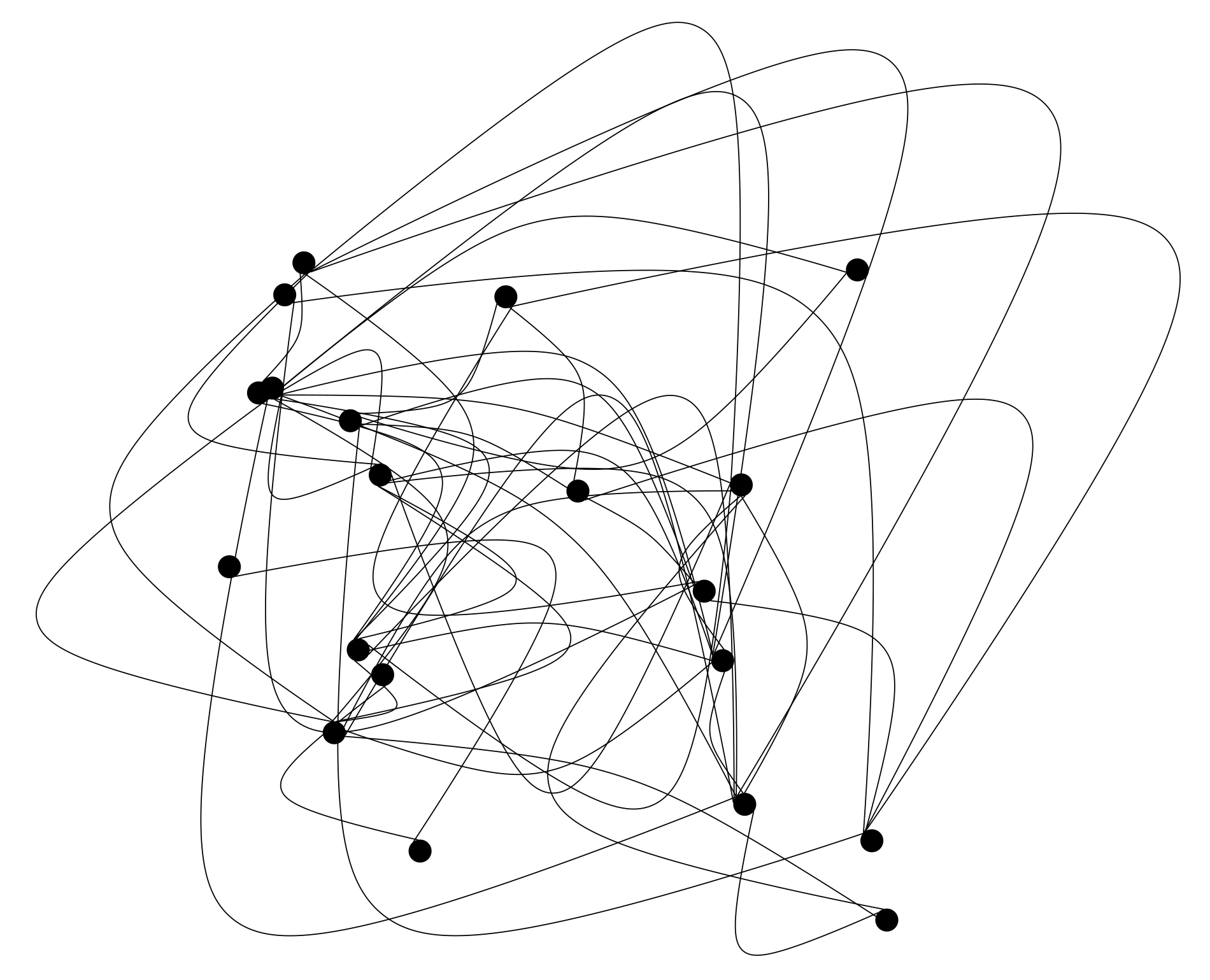}
}\hfill
\subfloat[Graph 2]{%
  \includegraphics[width=0.3\textwidth,keepaspectratio]{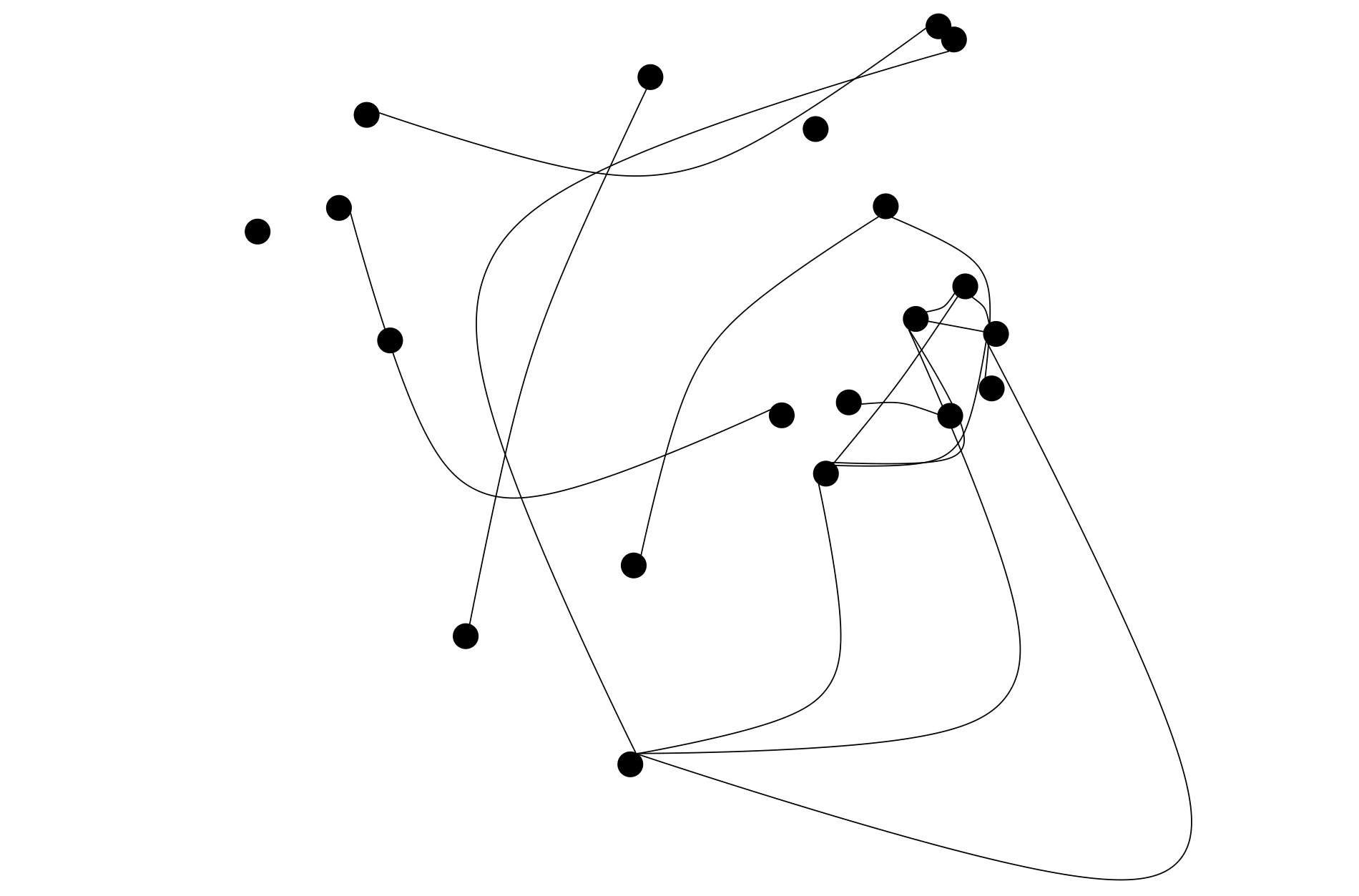}
}\hfill
\subfloat[Graph 3]{%
  \includegraphics[width=0.3\textwidth,keepaspectratio]{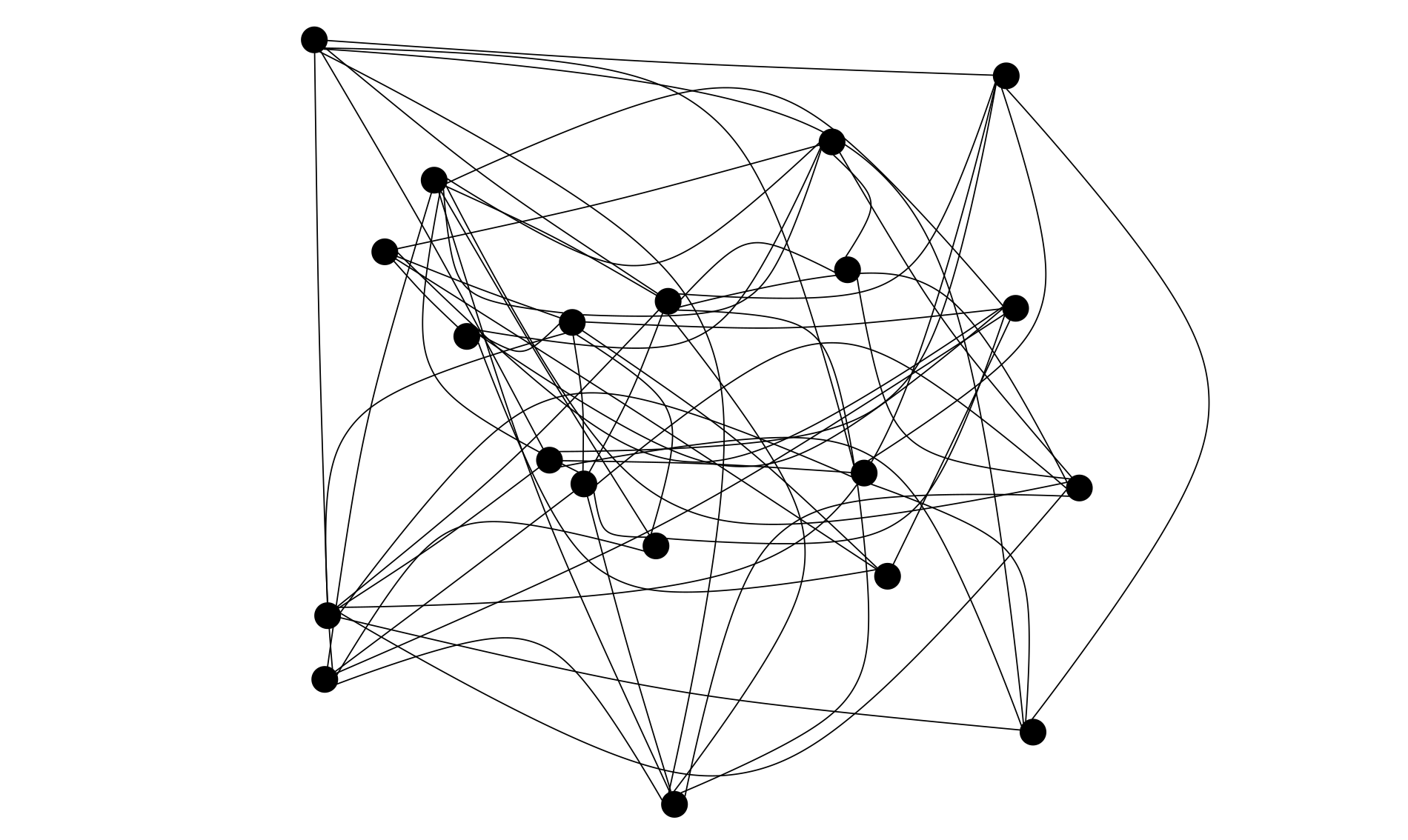}
}\\

\subfloat[Graph 4]{%
  \includegraphics[width=0.3\textwidth,keepaspectratio]{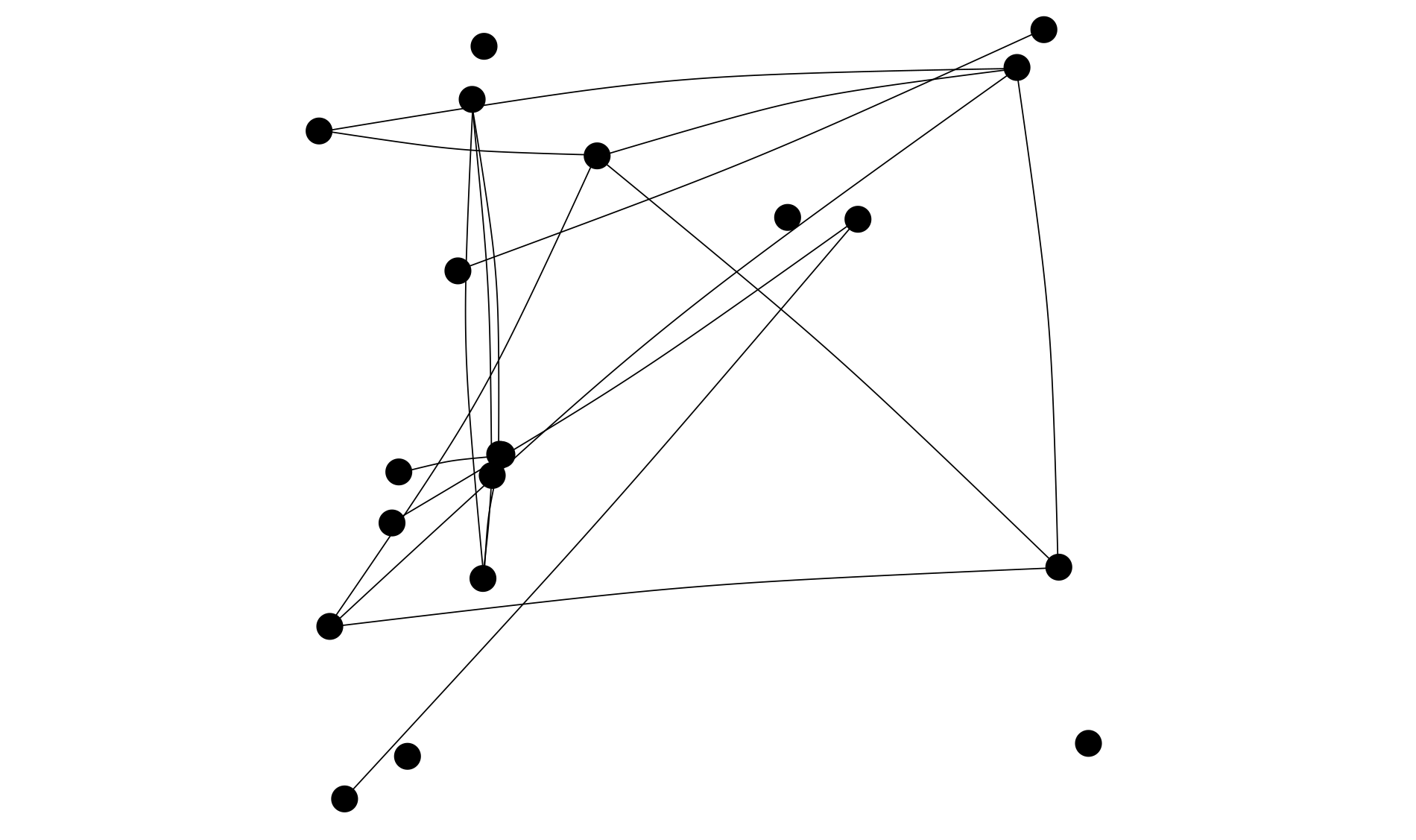}
}\hfill
\subfloat[Graph 5]{%
  \includegraphics[width=0.3\textwidth,keepaspectratio]{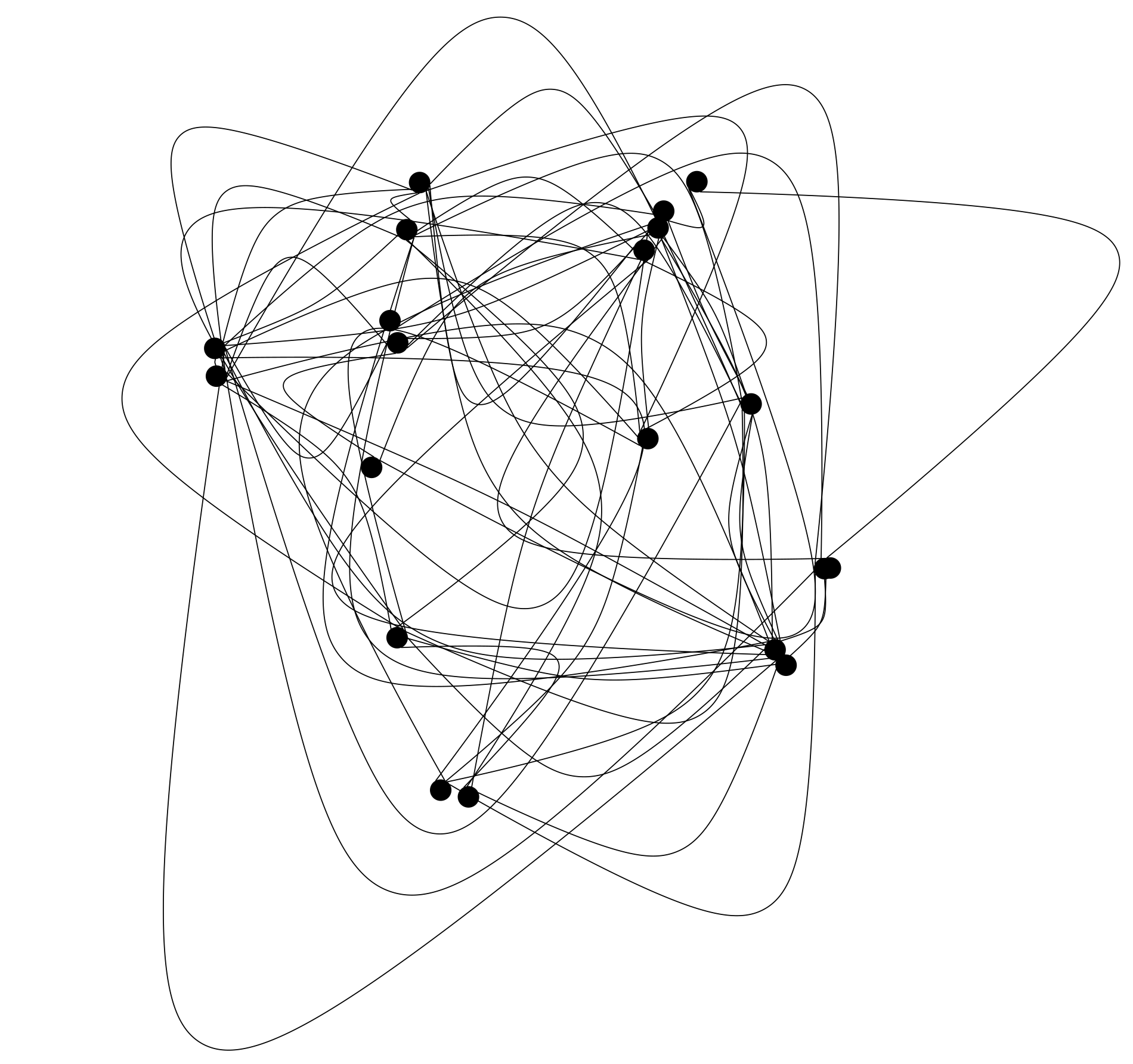}
}\hfill
\subfloat[Graph 6]{%
  \includegraphics[width=0.3\textwidth,keepaspectratio]{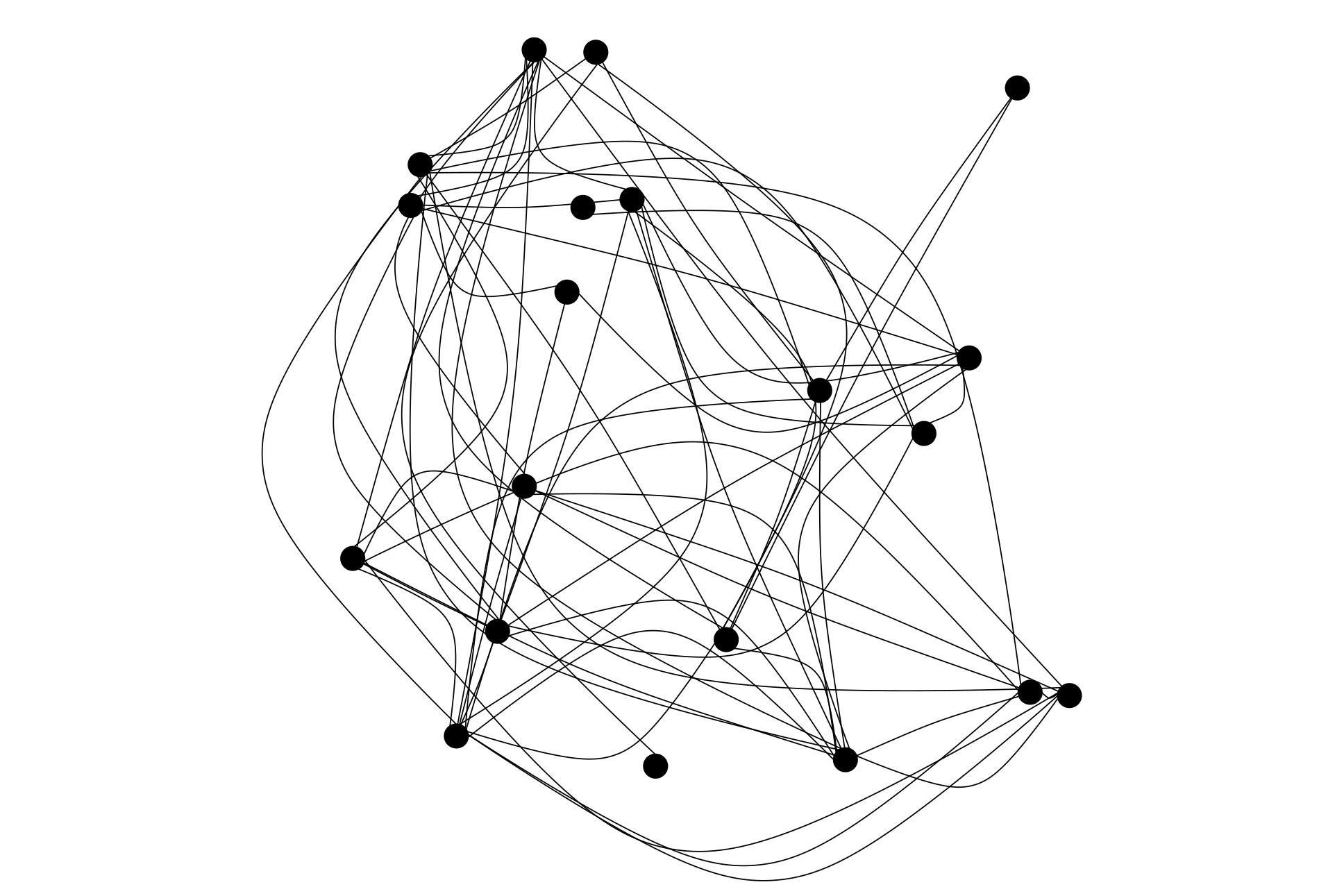}
}\\

\subfloat[Graph 7]{%
  \includegraphics[width=0.3\textwidth,keepaspectratio]{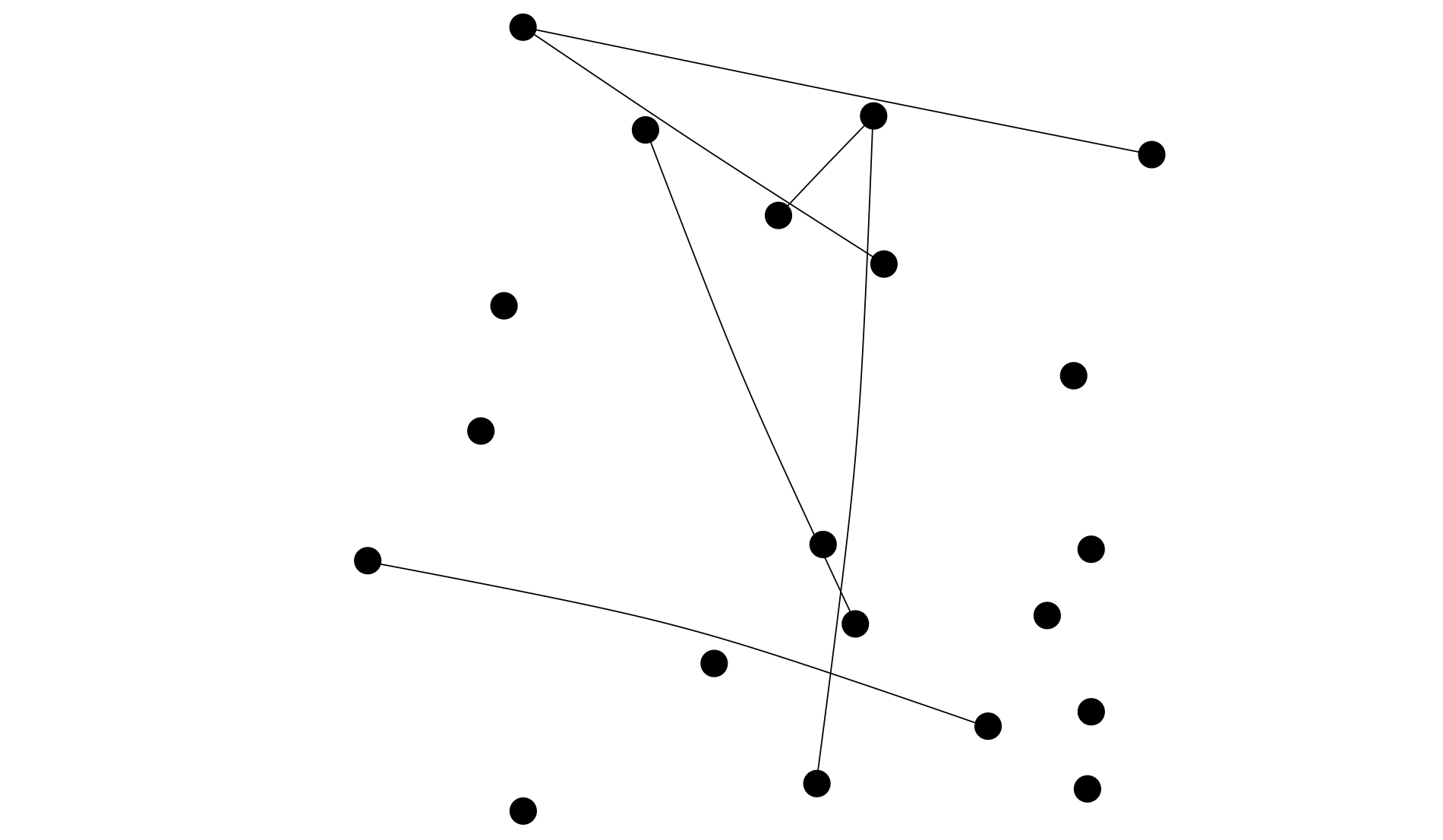}
}\hfill
\subfloat[Graph 8]{%
  \includegraphics[width=0.3\textwidth,keepaspectratio]{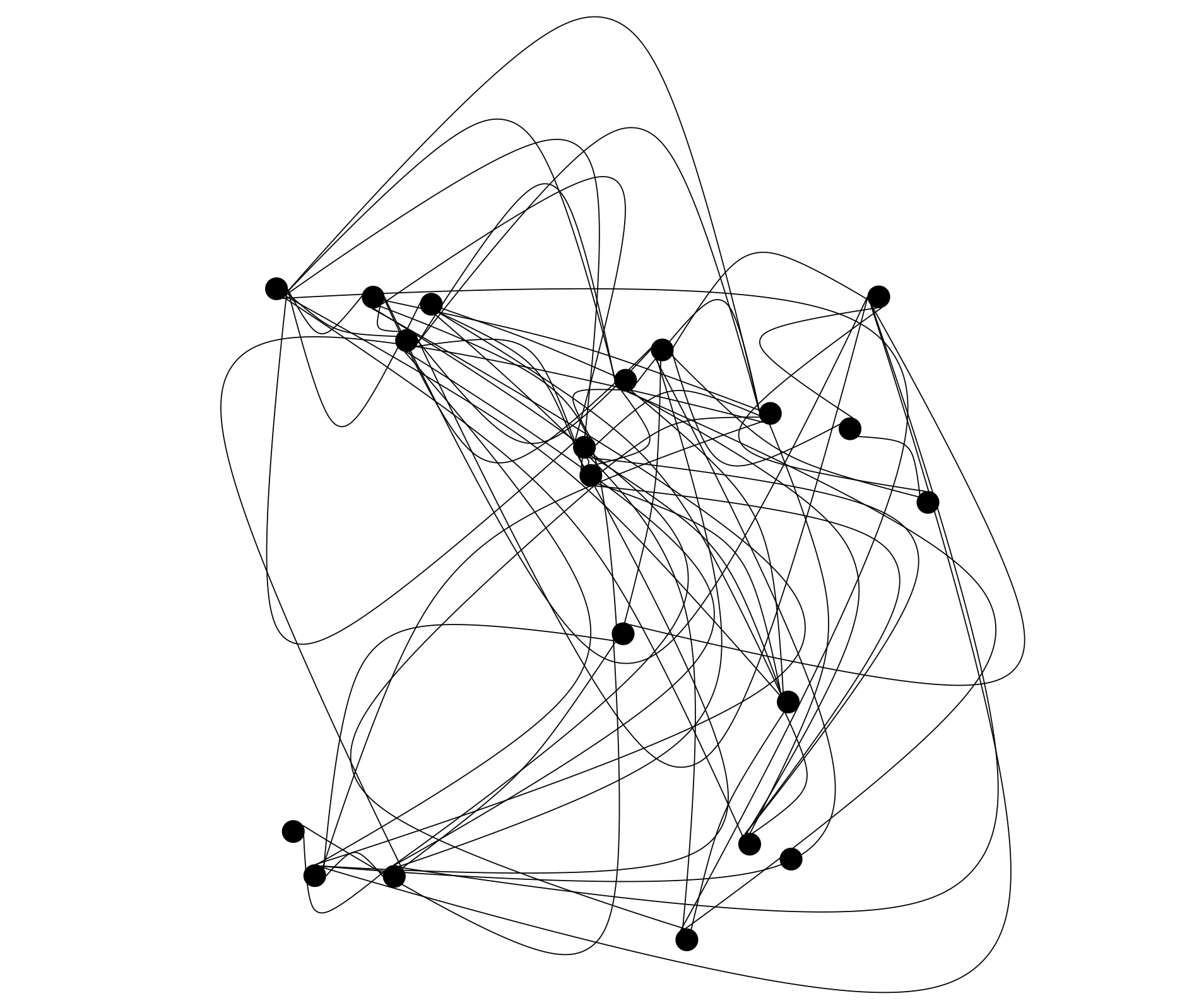}
}\hfill
\subfloat[Graph 9]{%
  \includegraphics[width=0.3\textwidth,keepaspectratio]{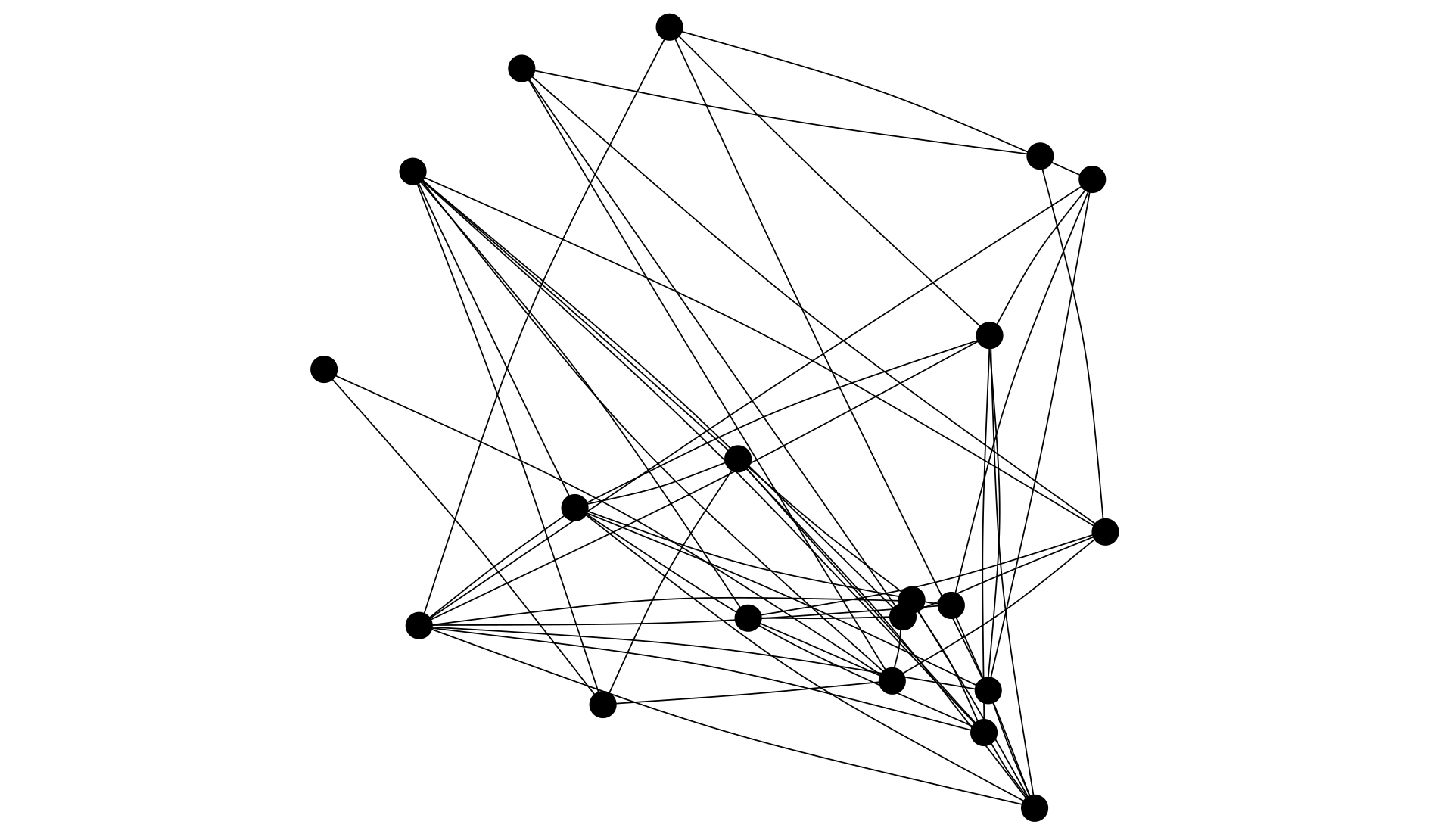}
}\\

\subfloat[Graph 10]{%
  \includegraphics[width=0.3\textwidth,keepaspectratio]{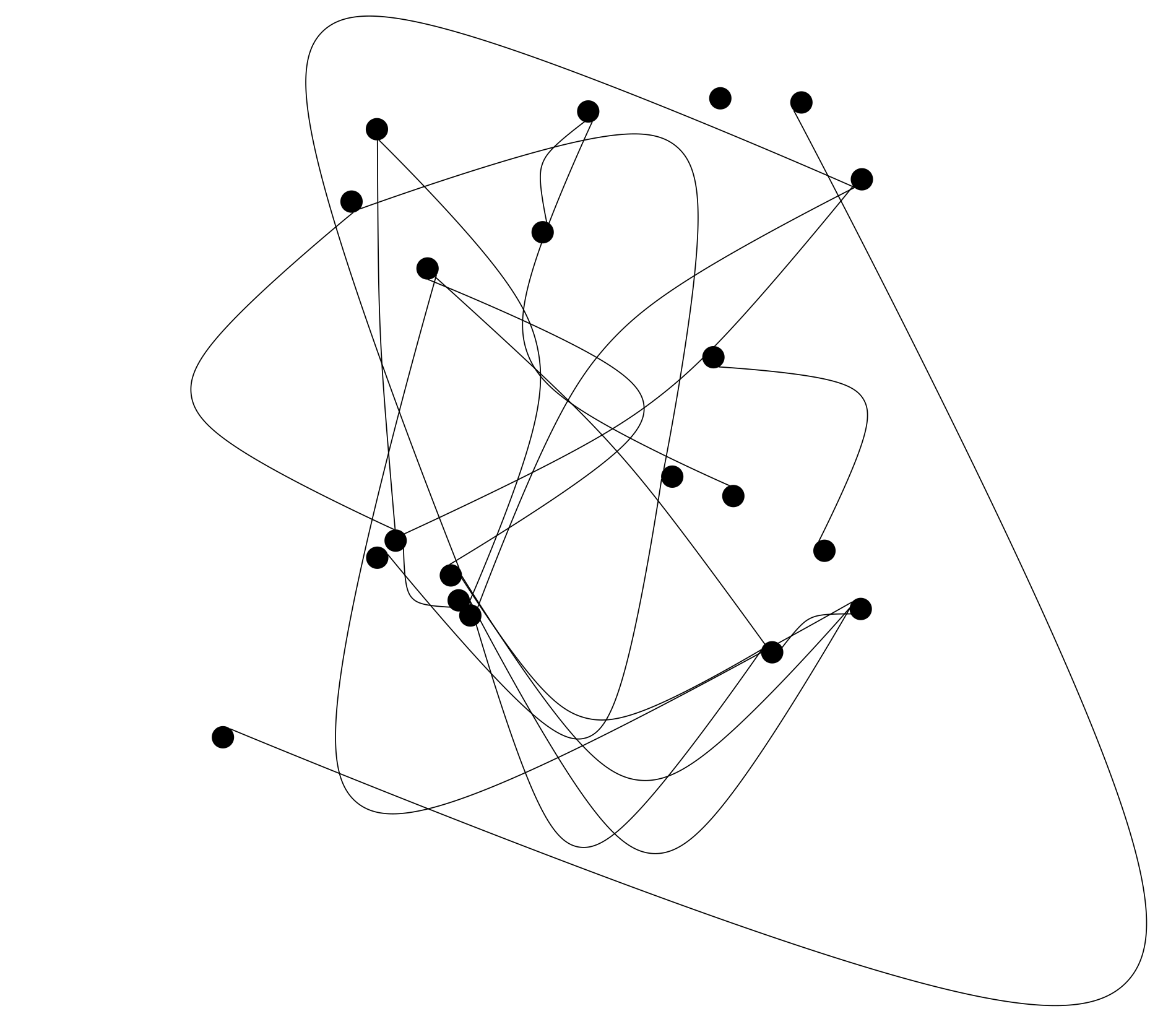}
}\hfill
\subfloat[Graph 11]{%
  \includegraphics[width=0.3\textwidth,keepaspectratio]{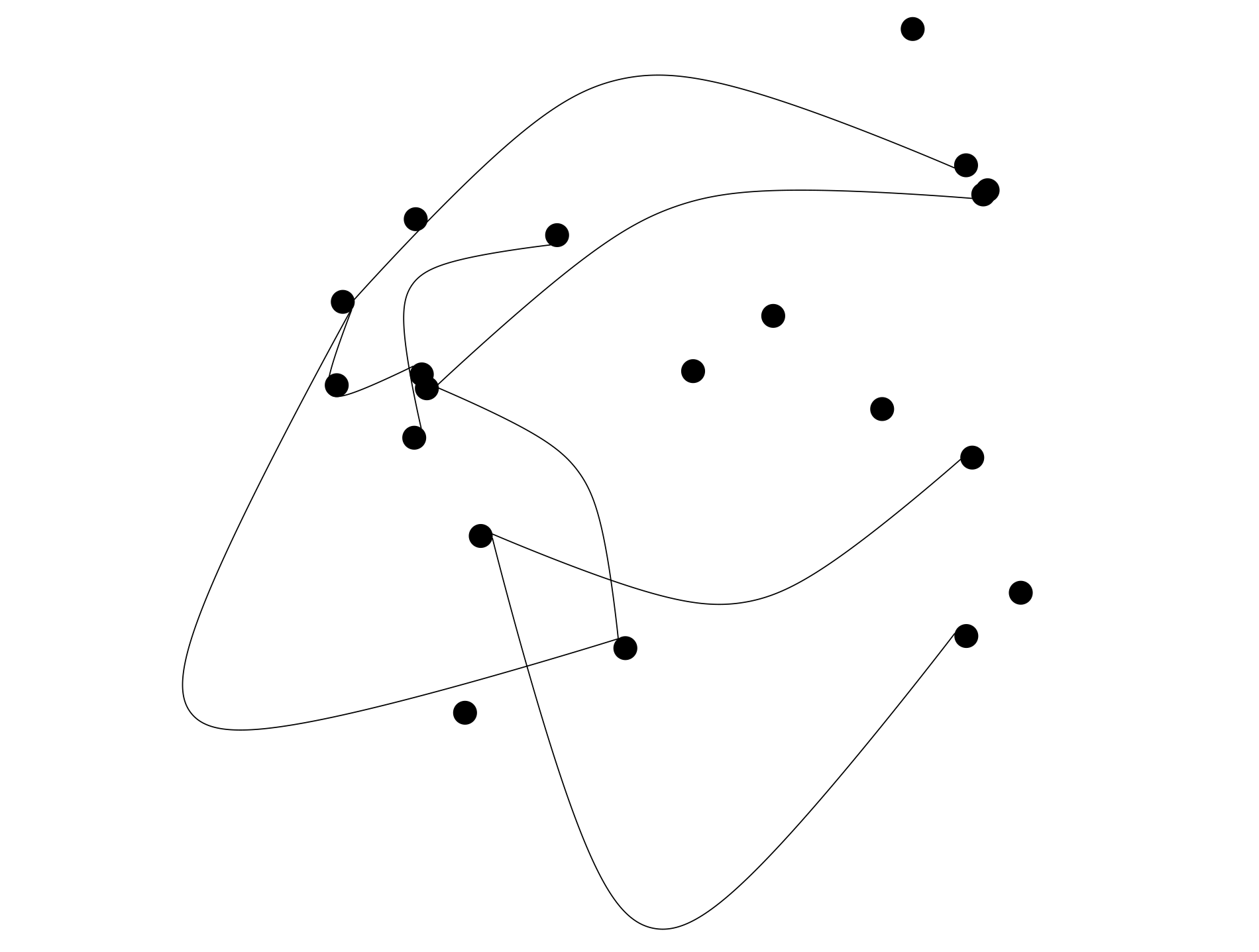}
}\hfill
\subfloat[Graph 12]{%
  \includegraphics[width=0.3\textwidth,keepaspectratio]{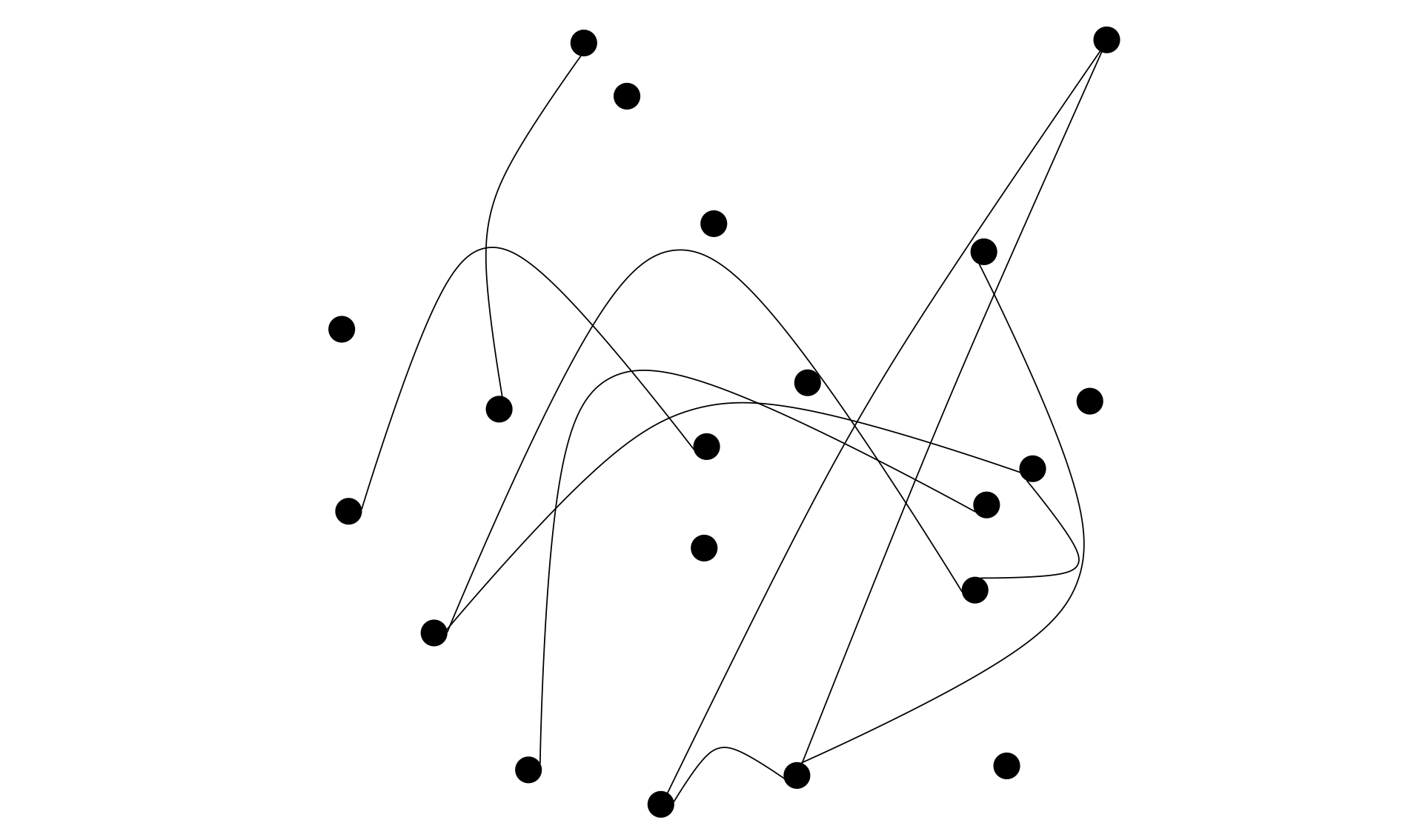}
}
\caption{Elements for Group B}
\end{figure}

\begin{figure}[!th]
\captionsetup[subfigure]{labelformat=empty}
\centering
\subfloat[Graph 1]{%
  \includegraphics[width=0.3\textwidth,keepaspectratio]{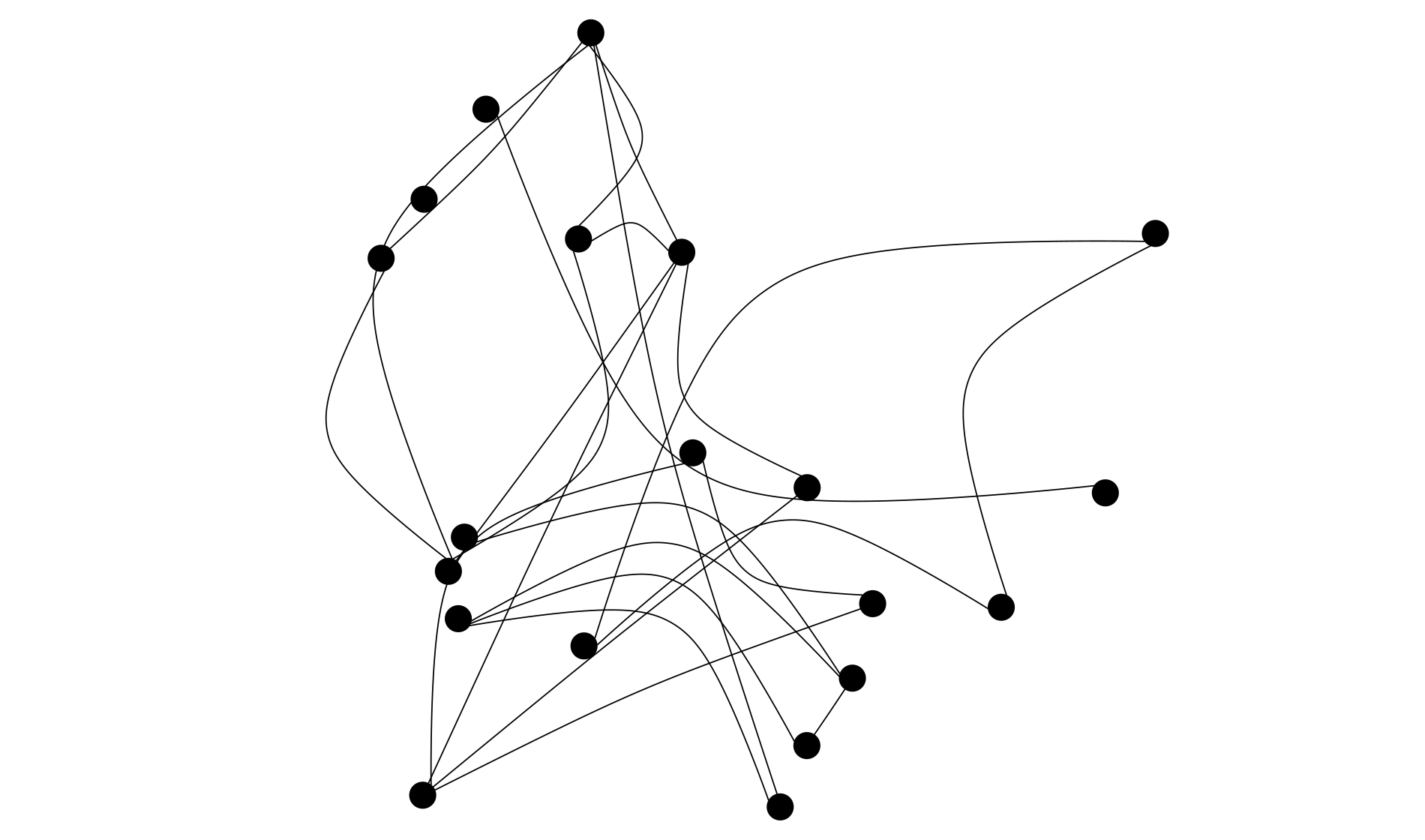}
}\hfill
\subfloat[Graph 2]{%
  \includegraphics[width=0.3\textwidth,keepaspectratio]{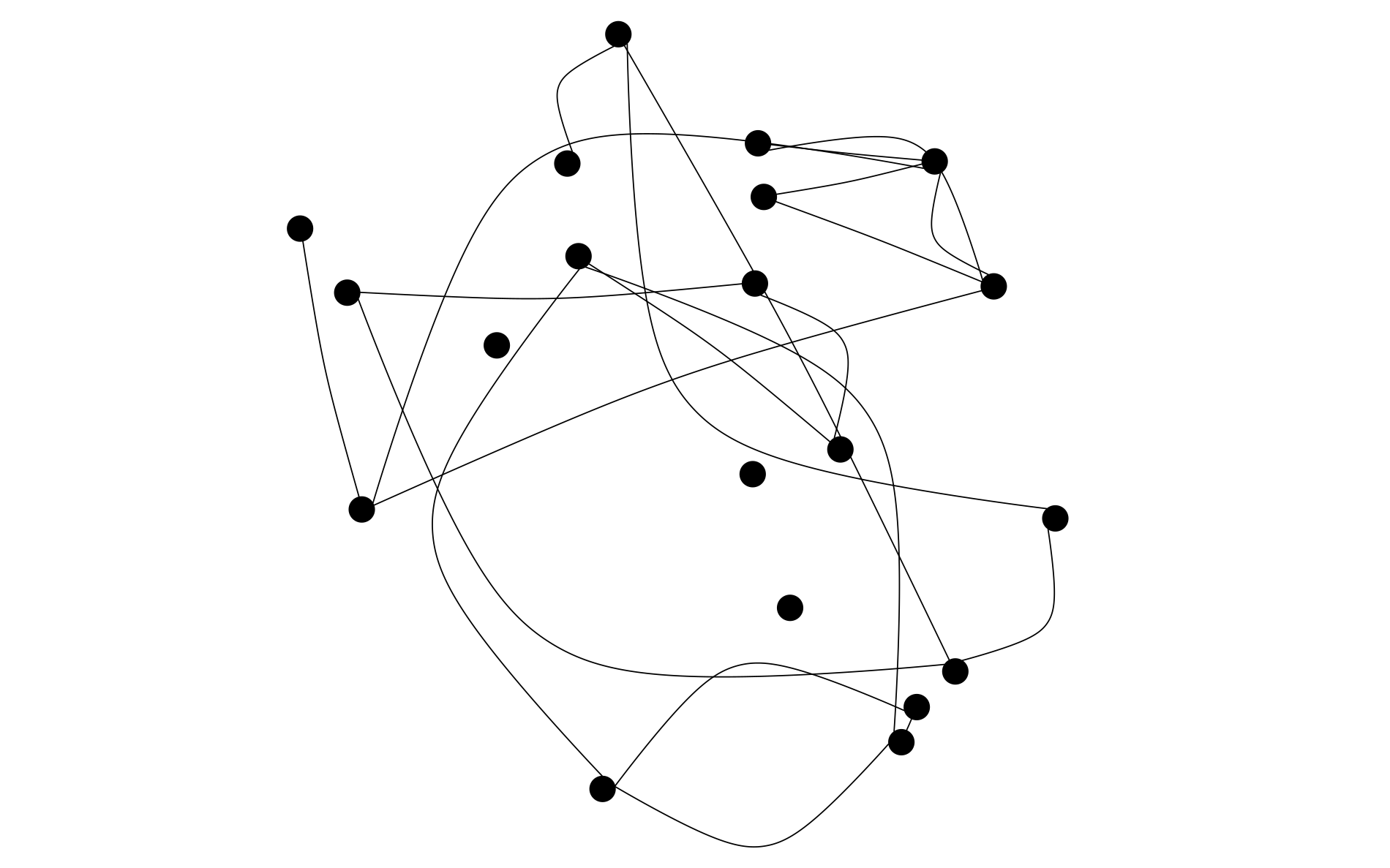}
}\hfill
\subfloat[Graph 3]{%
  \includegraphics[width=0.3\textwidth,keepaspectratio]{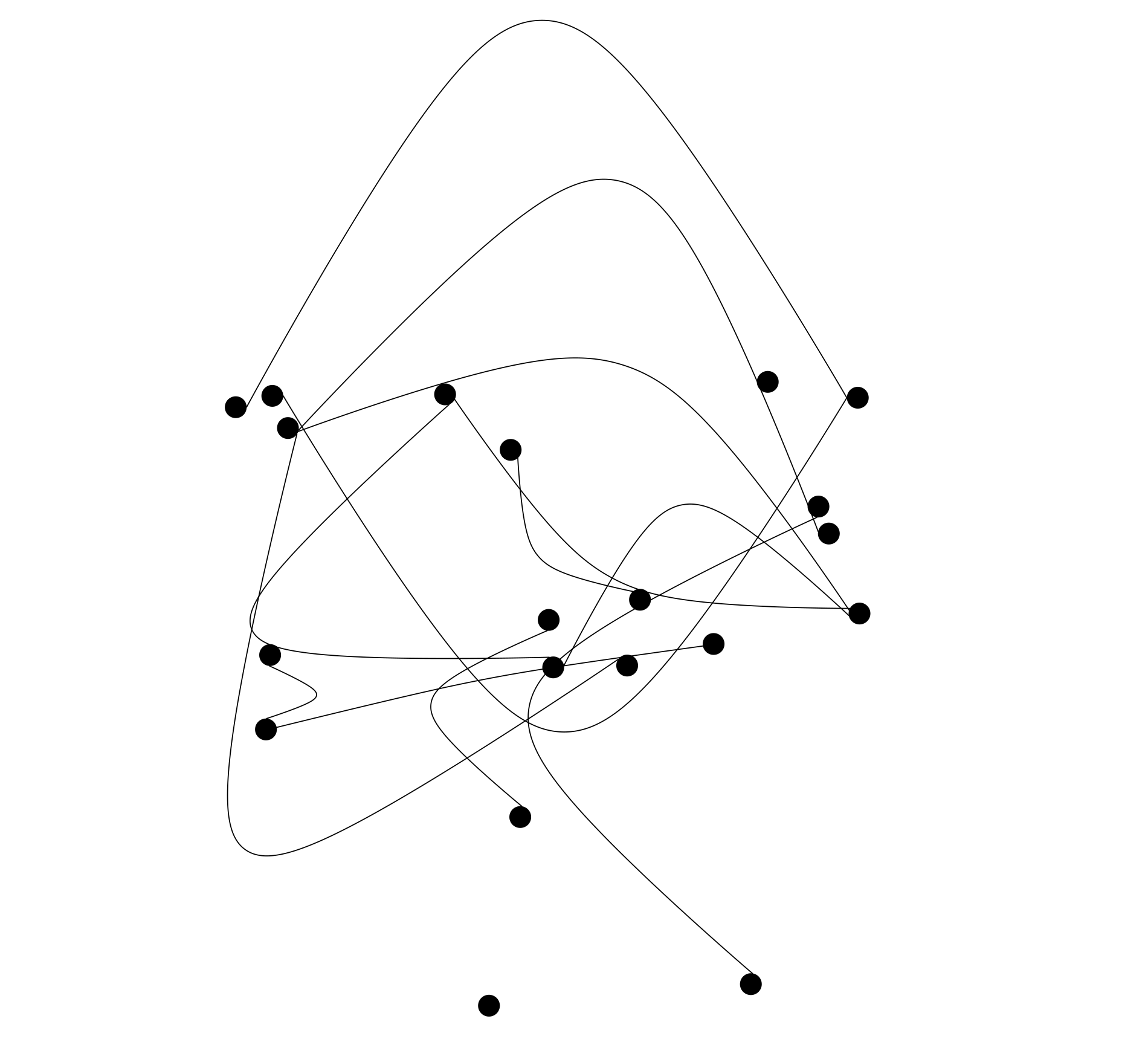}
}\\

\subfloat[Graph 4]{%
  \includegraphics[width=0.3\textwidth,keepaspectratio]{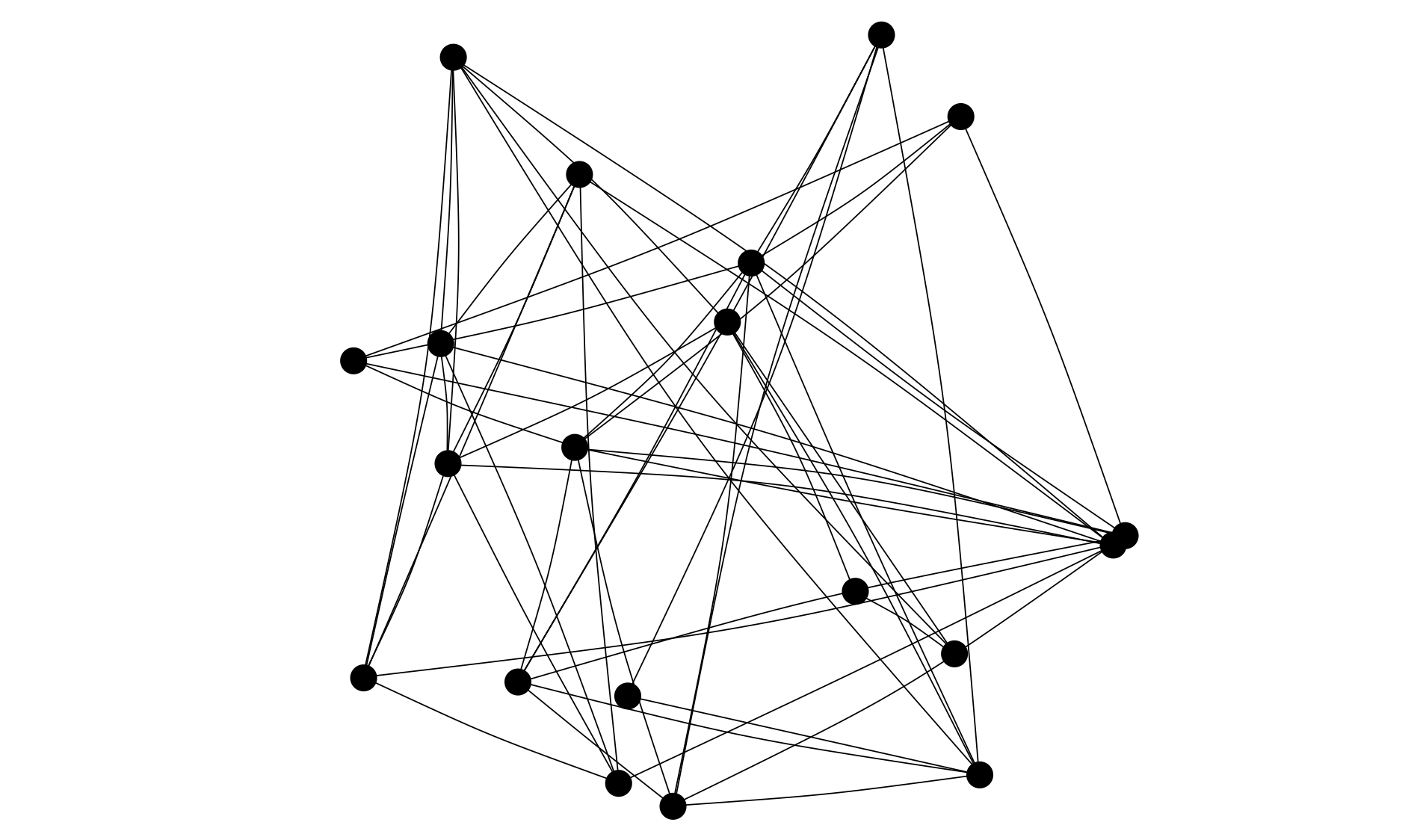}
}\hfill
\subfloat[Graph 5]{%
  \includegraphics[width=0.3\textwidth,keepaspectratio]{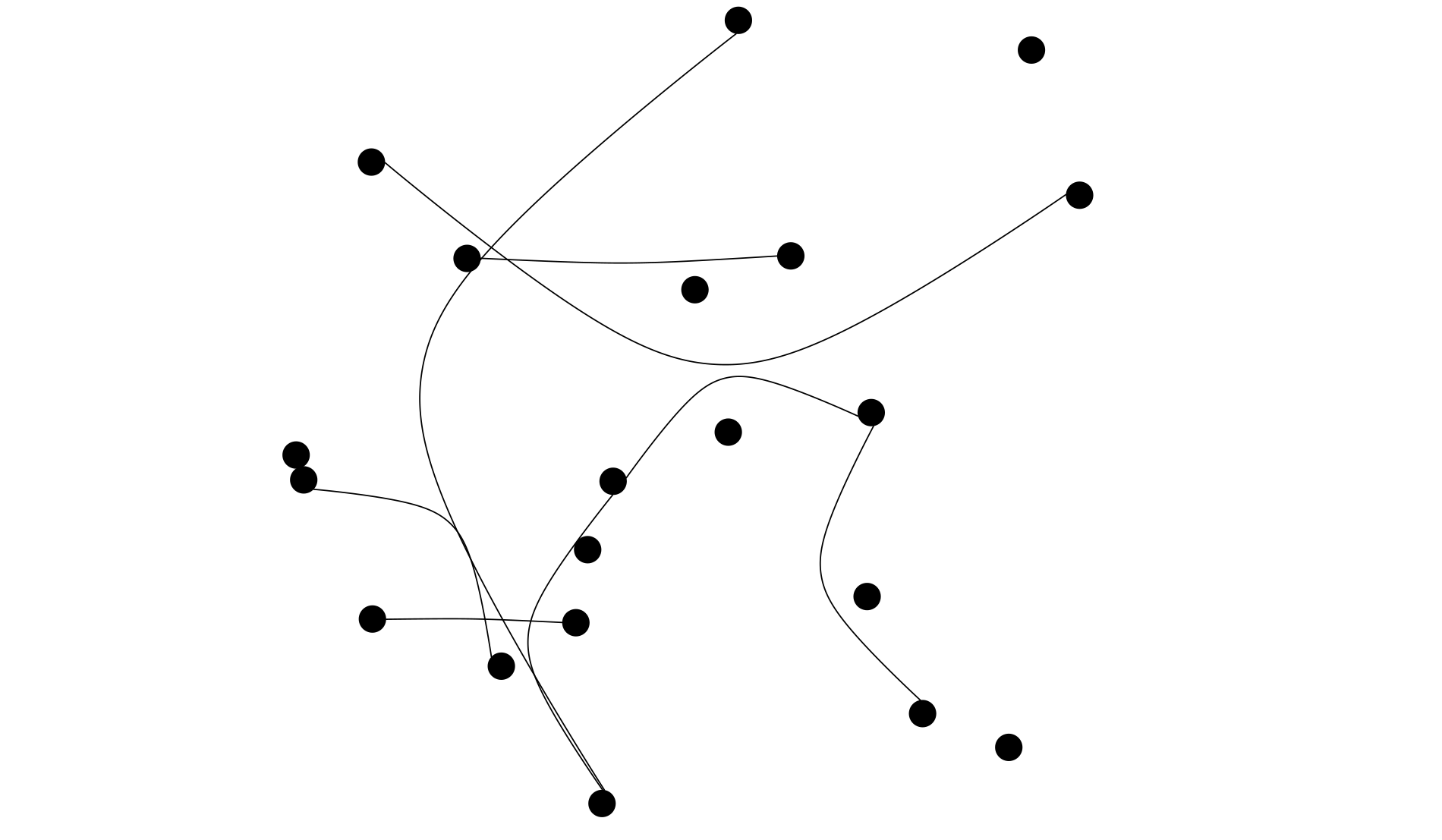}
}\hfill
\subfloat[Graph 6]{%
  \includegraphics[width=0.3\textwidth,keepaspectratio]{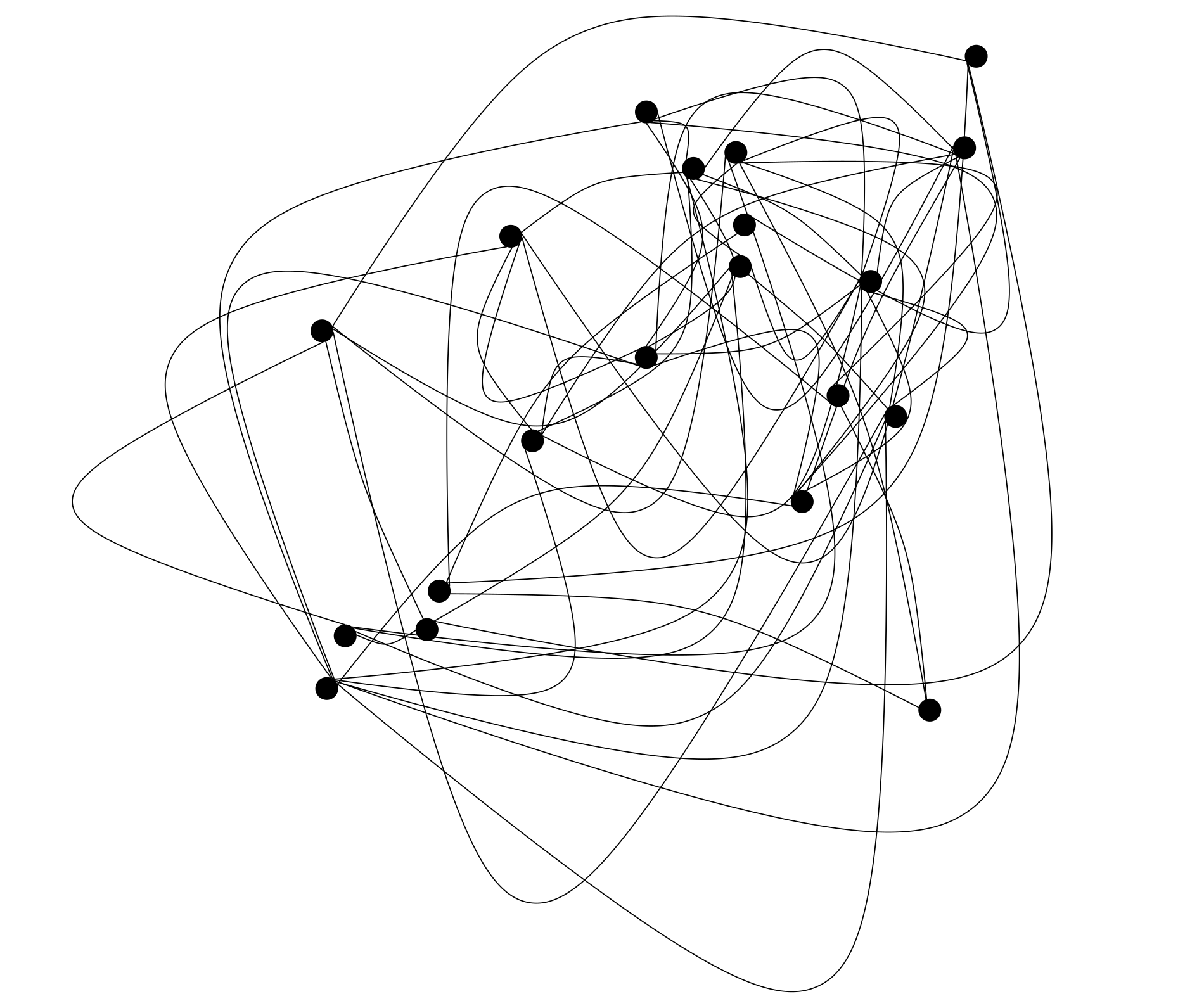}
}\\

\subfloat[Graph 7]{%
  \includegraphics[width=0.3\textwidth,keepaspectratio]{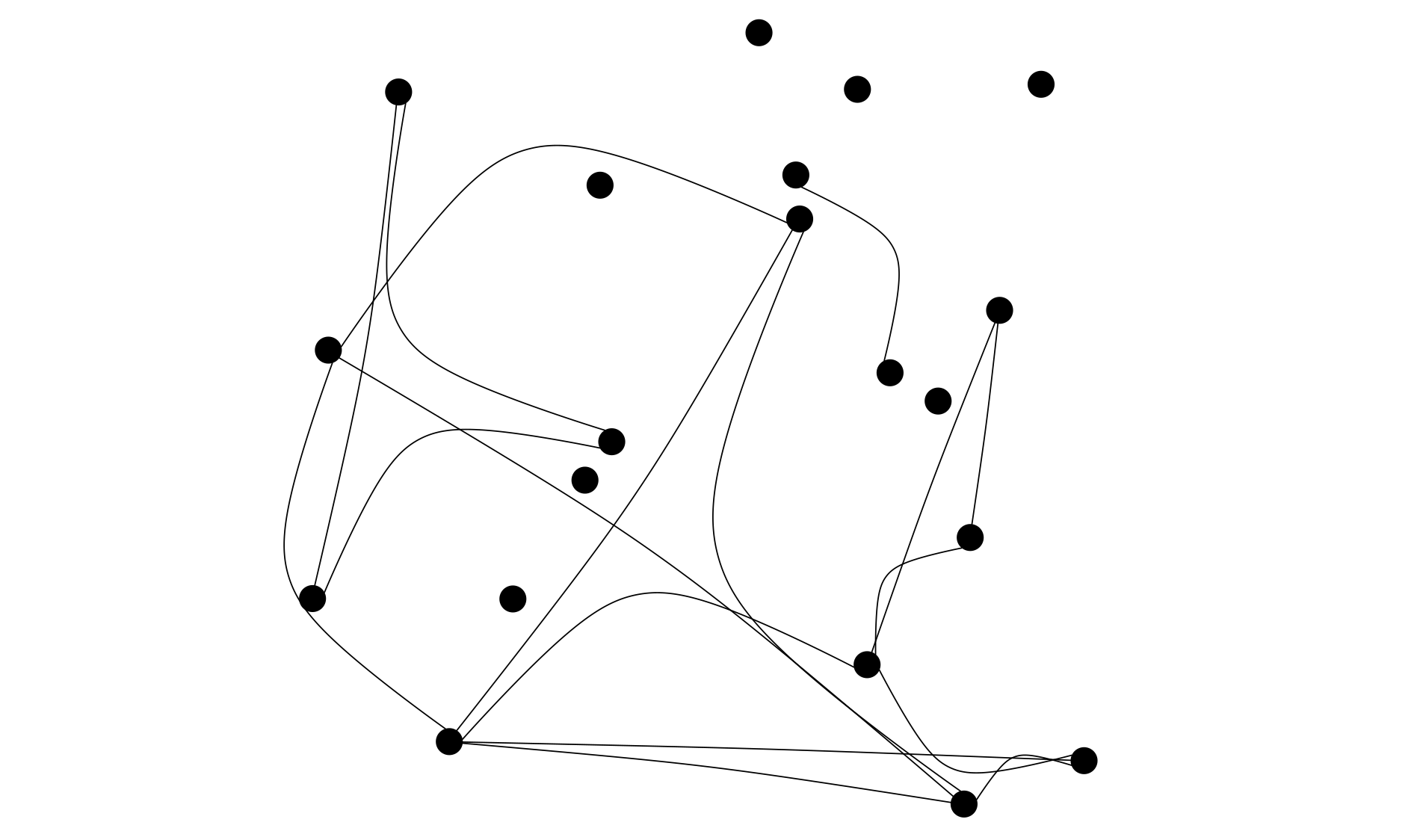}
}\hfill
\subfloat[Graph 8]{%
  \includegraphics[width=0.3\textwidth,keepaspectratio]{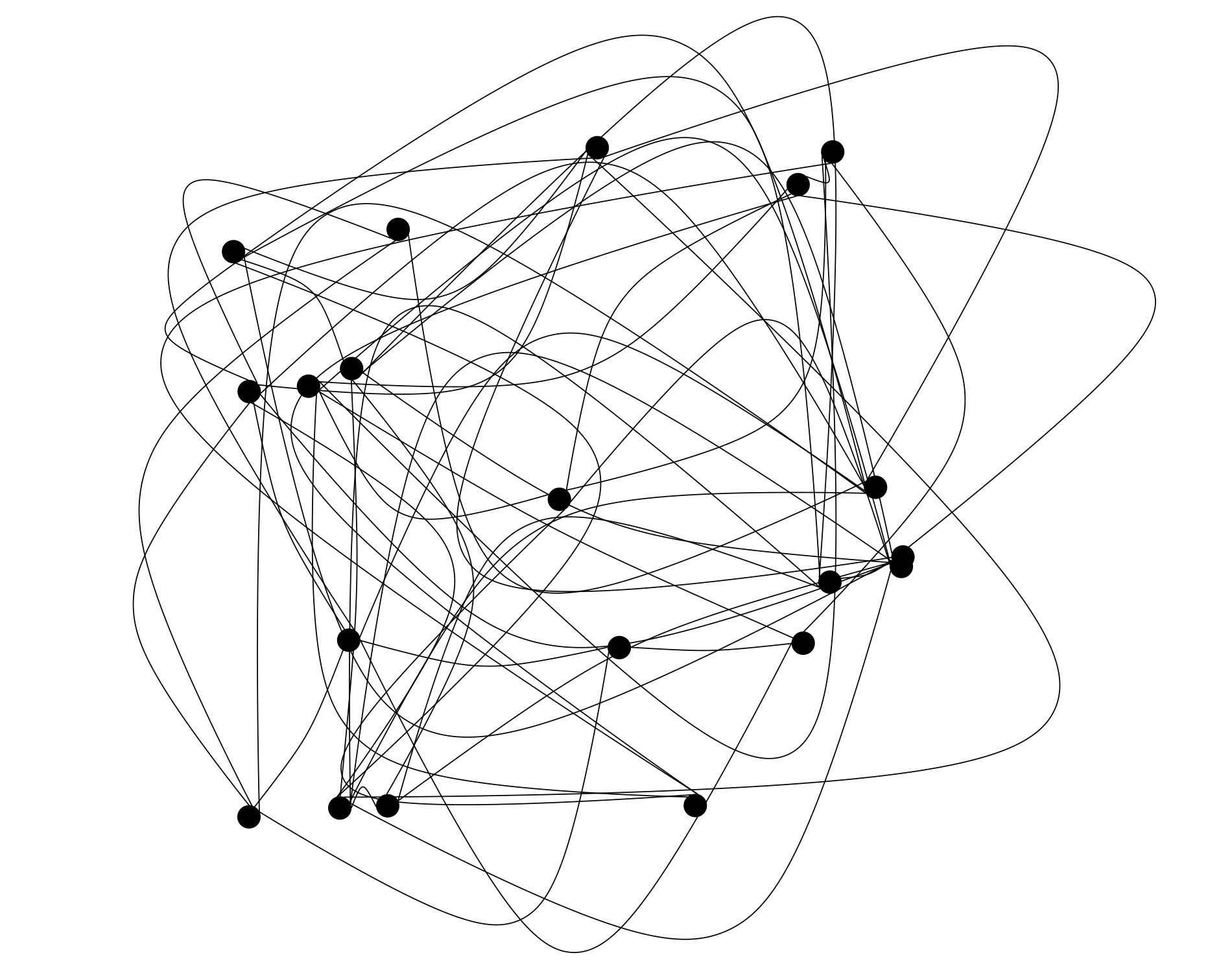}
}\hfill
\subfloat[Graph 9]{%
  \includegraphics[width=0.3\textwidth,keepaspectratio]{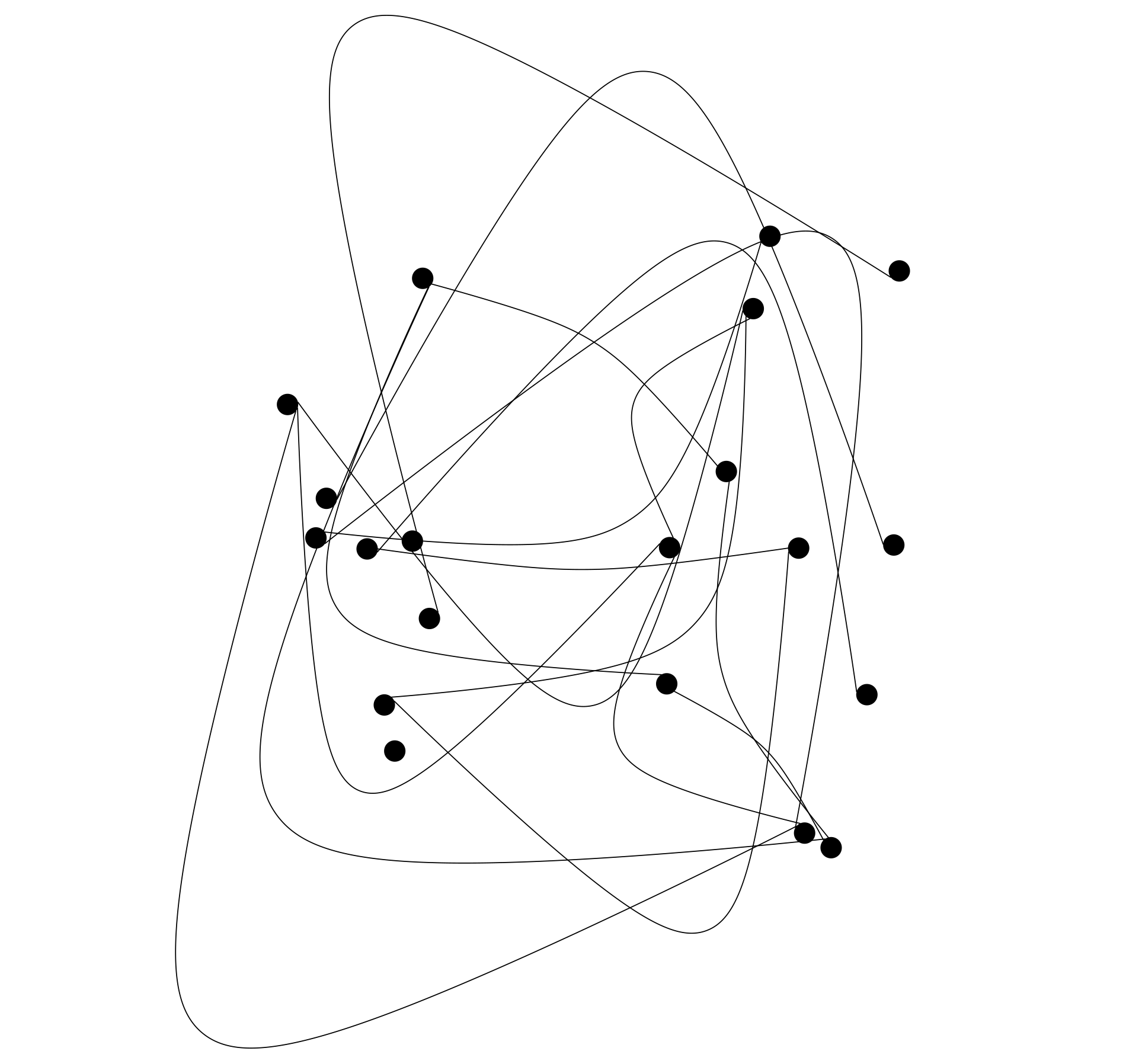}
}\\

\subfloat[Graph 10]{%
  \includegraphics[width=0.3\textwidth,keepaspectratio]{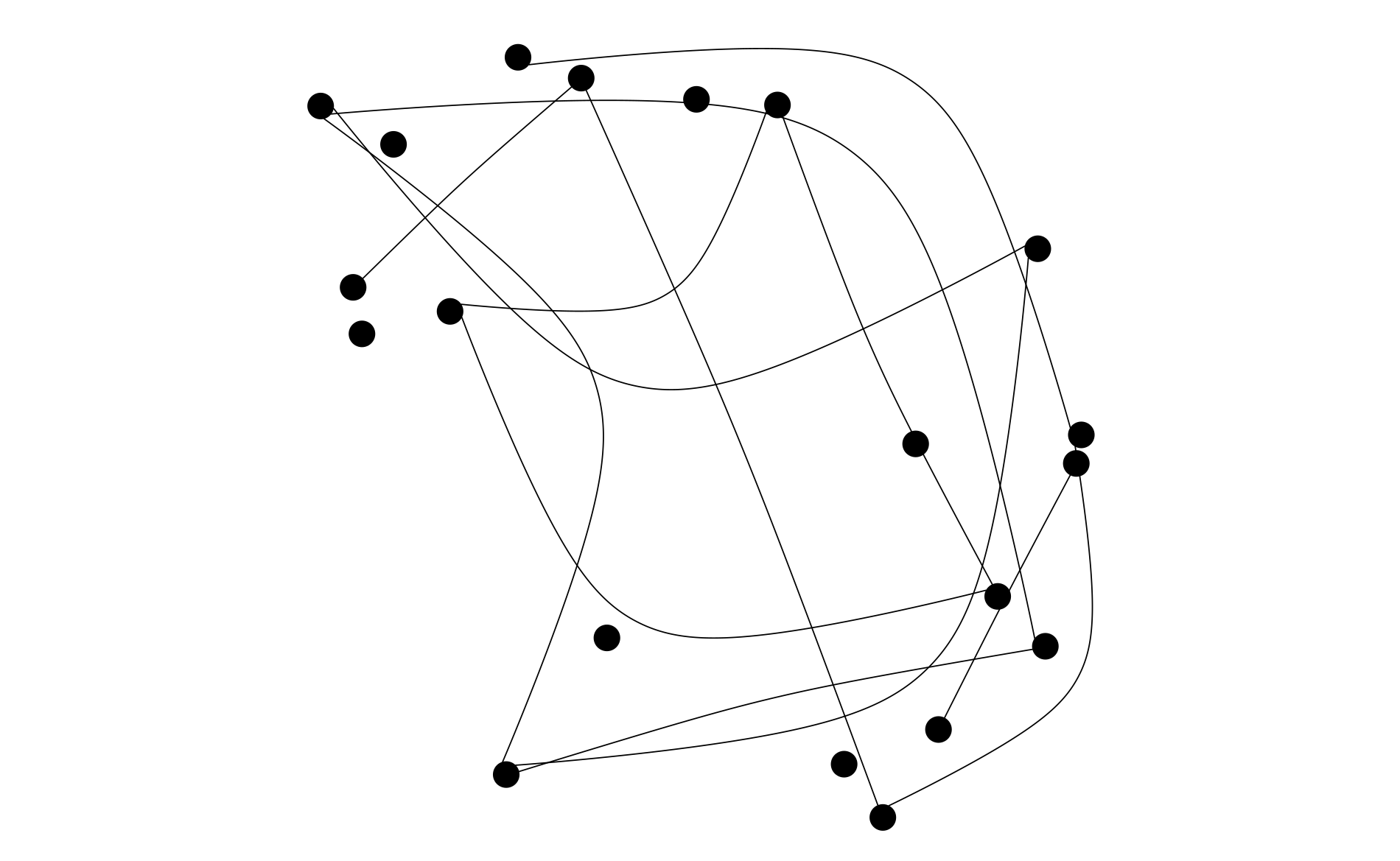}
}\hfill
\subfloat[Graph 11]{%
  \includegraphics[width=0.3\textwidth,keepaspectratio]{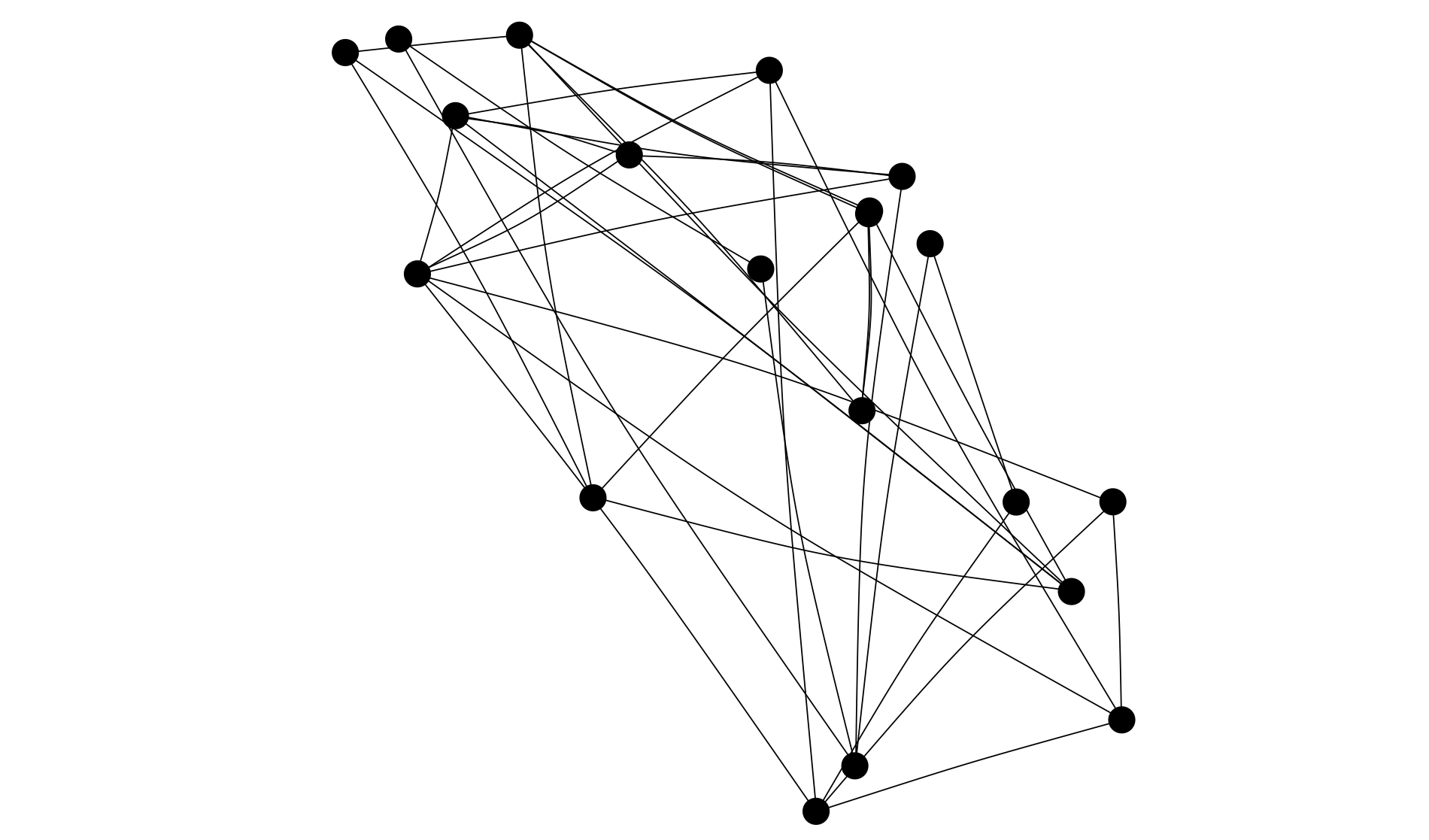}
}\hfill
\subfloat[Graph 12]{%
  \includegraphics[width=0.3\textwidth,keepaspectratio]{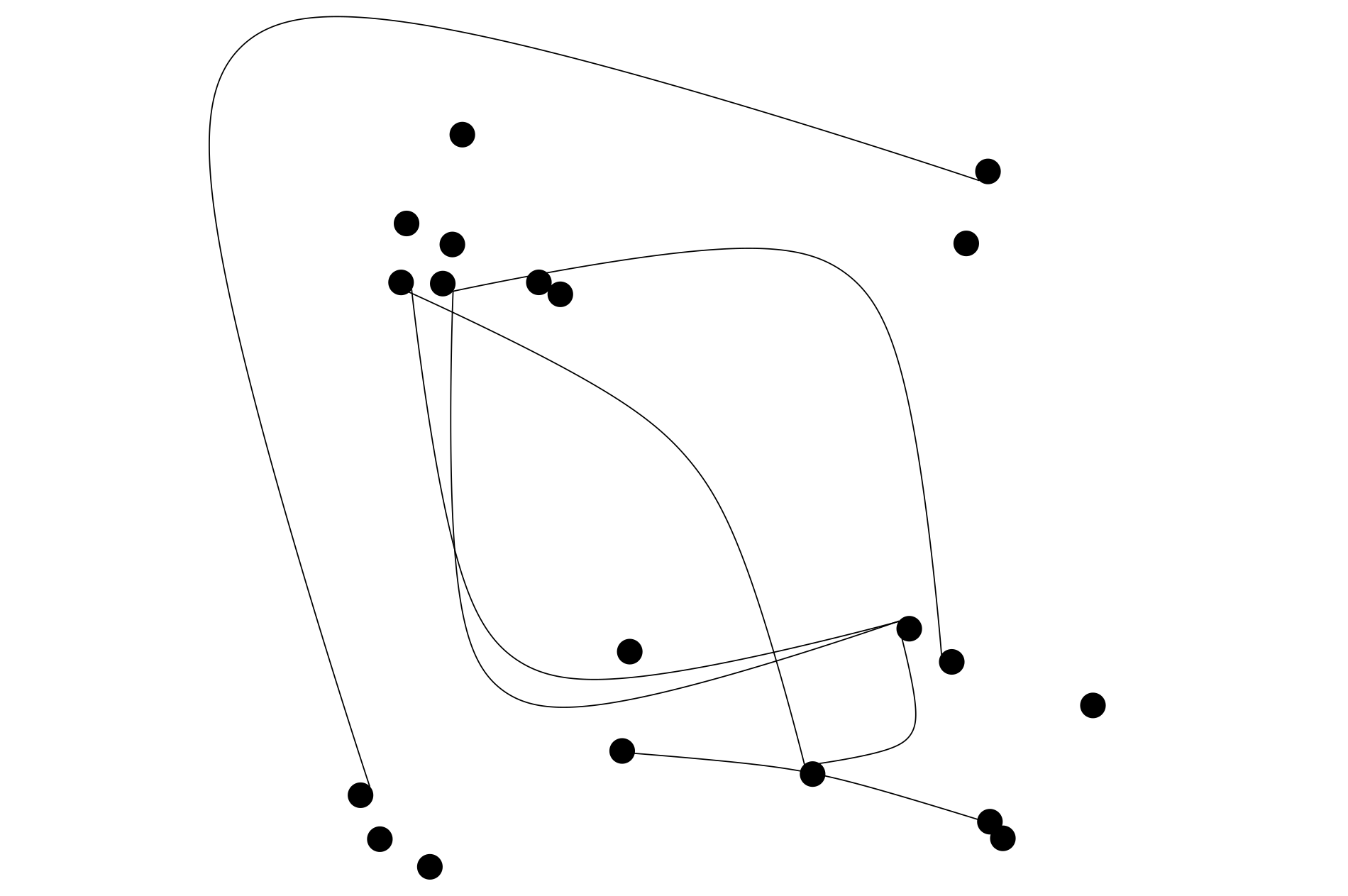}
}
\caption{Elements for Group C}
\end{figure}

\end{document}